# Fault diagnosis for PV arrays considering dust impact based on transformed graphical feature of characteristic curves and convolutional neural network with CBAM modules

Jiaqi Qu, Lu Wei[*], Qiang Sun, Hamidreza Zareipour, Zheng Qian[*]

**Author Names and Affiliations:**

Jiaqi Qu, Zheng Qian (Corresponding author): School of Instrumentation and Optoelectronic Engineering, Beihang University, Beijing, China.

Lu Wei (Corresponding author): School of Electronics and Information Engineering, Beihang University, Beijing, China.

Qiang Sun: State Key Laboratory of Control and Simulation of Power System and Generation Equipment, Department of Electrical Engineering, Tsinghua University, Beijing, China.

Hamidreza Zareipour: The Department of Electrical and Computer Engineering, University of Calgary, Calgary, Canada.

[*]**Corresponding Author:**

Lu Wei: School of Electronics and Information Engineering, Beihang University, Beijing, China.
E-mail: weilu@buaa.edu.cn
Zheng Qian: School of Instrumentation and Optoelectronic Engineering, Beihang University, Beijing, China.
E-mail: qianzheng@buaa.edu.cn
Tel: +86-010-82339267
Fax: +86-010-82339267

**Address:**

Xueyuan Road No.37, Haidian District, Beijing, 100191, China.

# Fault diagnosis for PV arrays considering dust impact based on transformed graphical feature of characteristic curves and convolutional neural network with CBAM modules

**Abstract:** Various faults can occur during the operation of PV arrays, and both the dust-affected operating conditions and various diode configurations make the faults more complicated. However, current methods for fault diagnosis based on I-V characteristic curves only utilize partial feature information and often rely on calibrating the field characteristic curves to standard test conditions (STC). It is difficult to apply it in practice and to accurately identify multiple complex faults with similarities in different blocking diodes configurations of PV arrays under the influence of dust. Therefore, a novel fault diagnosis method for PV arrays considering dust impact is proposed. In the preprocessing stage, the Isc-Voc normalized Gramian angular difference field (GADF) method is presented, which normalizes and transforms the resampled PV array characteristic curves from the field including I-V and P-V to obtain the transformed graphical feature matrices. Then, in the fault diagnosis stage, the model of convolutional neural network (CNN) with convolutional block attention modules (CBAM) is designed to extract fault differentiation information from the transformed graphical matrices containing full feature information and to classify faults. And different graphical feature transformation methods are compared through simulation cases, and different CNN-based classification methods are also analyzed. The results indicate that the developed method for PV arrays with different blocking diodes configurations under various operating conditions has high fault diagnosis accuracy and reliability.

**Keywords:** Photovoltaic; Fault diagnosis; Dust impact; Graphical feature transformation; Characteristic curves

## 1. Introduction

Solar energy, with its advantages of cleanness, accessibility, and utilization, has been paid increasing attention in the fight against global warming and fossil energy shortages. The International Energy Agency (IEA) predicts that the global PV market will grow by 25% year-on-year to 197 GW of new installed capacity in 2022, with cumulative installed capacity exceeding 1,000 GW. From 2022 to 2031, the global PV grid-connected installed capacity will grow at an average annual rate of 8% [1]. However, PV systems are often subject to abnormal failures due to disturbances in the operating environment, which can result in an estimated annual energy loss of up to 18.9% [2].

PV arrays, as the core component of PV systems, are prone to multiple faults such as short-circuit, open-circuit, abnormal degradation, partial shading, etc. in complex outdoor environments [3]. Currently, the traditional DC-side protection methods, such as over-current protection (OCPDs) and ground fault protection (GFPDs), have certain difficulty in determining the type of faults [4, 5]. Furthermore, due to the non-linear output characteristics of the PV array, both low mismatch faults and faults at low irradiance may cause the failures of the protection [6]. These failures can not only affect power generation but also lead to serious safety issues. In some cases, faults may even increase the risk of fire and personal danger, if not detected and corrected in time [7]. In addition, the dusty environment has a significant impact on the operation performance of PV systems [8-10]. Dust from PV panels can reduce the power of PV systems [11], and more importantly, the long-term dust deposition operating conditions also complicate faults, forming compound faults that are more difficult to classify [12]. Therefore, effective fault detection and diagnosis of PV arrays under the influence of dust is essential for the safety and reliability of PV systems [13].

At present, there are two main categories of techniques for detecting and diagnosing faults in PV arrays. One is the offline diagnosis methods, which include earth capacitance measurement (ECM) [14] and time domain reflection (TDR) [15, 16], etc. These methods rely on specific external signal generators for testing and require the offline operation of the PV array, which interferes with normal operation and makes it hard to diagnose faults online in real-time. The other category is the online diagnostic methods that do not affect the normal operation of the PV system. According to the diagnostic indicators, these online methods are further divided into those based on the measurement of current and voltage output on the DC side of each substring of the array [4, 17, 18], those based on the measurement of current and voltage output on the AC side of the array [6, 19, 20], and those based on the I-V characteristic curves [21-23]. Specifically, the DC indicators-based method analyses faults through the output residuals of the current or voltage in each substring relying on a large number of sensors, which is costly and has limited universality. The AC indicators-based method does not need to install a large number of sensors and uses the residual thresholds of current and voltage output on the AC side of the array. However, the types of faults that can be distinguished between different threshold intervals are inadequate, and it is hard to diagnose multiple complex faults with complete accuracy. Since I-V characteristic curves usually contain rich information on the status of PV modules, diagnosis based on I-V curves is a hot topic [24]. Moreover, the I-V tracer currently already supports measurements for a single module or small-scale strings or arrays and has realized online measurement without changing the operational state [25, 26]. In this sense, the diagnosis method based on I–V curves can be applied to all common PV installations and is easier to implement in the field.

In existing studies, according to forms of curve features applied, fault diagnosis methods using I-V characteristic curves can generally be divided into: i) raw I-V curves relying on deep learning models for diagnosis and ii) extracted key features of I-V curves for diagnosis. Further, the former includes: (1) taking I-V curves data as input directly. For example, Chen et al. [27] assembled I-V curves with irradiance ($G$) and module temperature ($T$) into a 4-column matrix to classify 8 classes of PV array faults by an improved ResNet model. Gao et al. [28] designed a fusion model of convolutional neural network (CNN) and residual-gated recurrent unit (Res-GRU) to diagnose hybrid faults using a combination matrix of I-V curves, irradiance and temperature as inputs. (2) taking the residual between the measured I-V curve and the theoretical curve as input. For example, Chine et al. [29] compared the difference between the measured and the simulated PV array output power and identified faults by the attributes of differences in I-V curves. Liu et al. [21] proposed a fault diagnosis method based on stacked auto encoder (SAE) and clustering, which extracts features from the difference between the simulated and measured I-V curves to achieve classification. These methods have insufficient ability to process the full feature information contained in the original I-V curves completely, and the extraction of features mostly depends on classifiers with complex structures. In addition, they only enable the classification of a fixed degree of faults and have tiny diagnostic ability for the full fault levels of defects. The latter includes (3) identifying key features of array characteristics (e.g. $V_{OC}$, $I_{SC}$, $V_{MPP}$, $I_{MPP}$, $FF$, $R_S$, and $R_P$) in the curves as input. For example, Fadhel et al. [30] adopt $V_{MPP}$, $I_{MPP}$, and $P_{MPP}$ as features to classify four different shading configurations. Liu et al. [31] extracted five key points from I-V curves as valid features as input to a fault diagnosis method based on variable prediction model. Besides, some similar approaches are presented in [32-34]. (4) calculating shape features (e.g. derivative and curvature) of the curves as input. For example, Bressan et al. [35] proposed method based on the analysis of the first and second derivative of the I-V curves, for detecting faults on series resistors and activation of bypass diodes. Ma et al. [36] analyzed the extraction of negative peaks on the derivative of the I-V curves, whereby single faults and compound faults at different levels of shading were diagnosed. However, it should be pointed out that these studies only used part information of characteristic curves. The diagnosis was completed by analyzing the current ($I_{MPP}$), voltage ($V_{MPP}$), power ($P_{MPP}$) or curve shape features at the maximum power point (MPP). In some complex scenarios such as faults considering soiling, the feature extraction process is complicated, and they may appear to have the same MPPs leading to wrong classifications.

Recently, several studies have proposed additional contributions to this field. Lin et al. [37] extracted multi-scale fault features using different scales of convolution horizons and identified fixed fault parameters of faults effected by soiling, Huang et al. [38] investigated full-scale faults under the operating condition with soiling impact in PV arrays without blocking diodes, and Li et al. [39] used full characteristic information of the graphical features of I-V curves and machine learning techniques for PV arrays fault diagnosis. However, several gaps still remain. First, to the best of the authors' knowledge, there are no studies that provide a comprehensive analysis of full-scale faults with dynamic fault parameters in PV arrays with various blocking diode configurations considering the impact of soiling. Second, extracting features of array characteristics or curve shapes does not make effective use of full information contained in the I-V curves, leaving incomplete types of PV array faults to be identified. Furthermore, most current diagnostic methods using characteristic curves rely on additional measured curves of fault samples to calibrate the I-V curves to the standard test conditions, severely restricting their practical applicability. Therefore, this paper aims to develop a feature processing method using full information of characteristic curves, which does not require additional experiments to calibrate curves under different environments. Furthermore, a full-range multi-faults classification method adapted to different PV array configurations considering soiling impact is designed. It enables the complete diagnosis of complex faults for various operating conditions and configurations of PV arrays. The main contributions of the method proposed in this paper are:

(1) Various fault types of PV arrays in multiple scenarios are compared and analyzed, including PV arrays with and without blocking diode configurations and compound fault types with and without the influence of soiling. It overcomes the challenge of uniqueness of array fault diagnosis methods for different objects and provides universal ability to diagnose different PV configurations with dynamic array faults.

(2) The graphical feature transformation method based on Isc-Voc normalized GADF is proposed to extract common features of the same fault in different environments, and the transformed graphical features of the characteristic curves are stacked into 2-channel 2D matrices as input features for the classification model. It fills the gap of diagnosing complex faults affected by soiling using the complete characteristic curves information without calibration experiments, which greatly improves the practical application value.

(3) The classification model of convolutional neural network with CBAM module is designed for fault diagnosis in multiple scenarios, which can accurately classify and identify the full range of complicated faults. It extends the performance of diagnosis methods under the influence of dust, and improves the diagnosis accuracy and robustness under multiple scenes.

The rest of this paper is organized as follows: Section 2 formulates the preliminary analysis of the problem to be studied. Section 3 details the proposed fault diagnosis methodology. Section 4 presents experimental results and discussion. Finally, Section 5 concludes this work.

## 2. Preliminary analysis

In this section, first, the fault behaviors of PV arrays with different blocking diode configurations are analyzed. Second, the characteristic curves under normal operating condition and condition considering the influence of soiling are compared. Furthermore, the preprocessing techniques used for the current fault diagnosis are investigated, and the limitations of the existing preprocessing methods are identified.

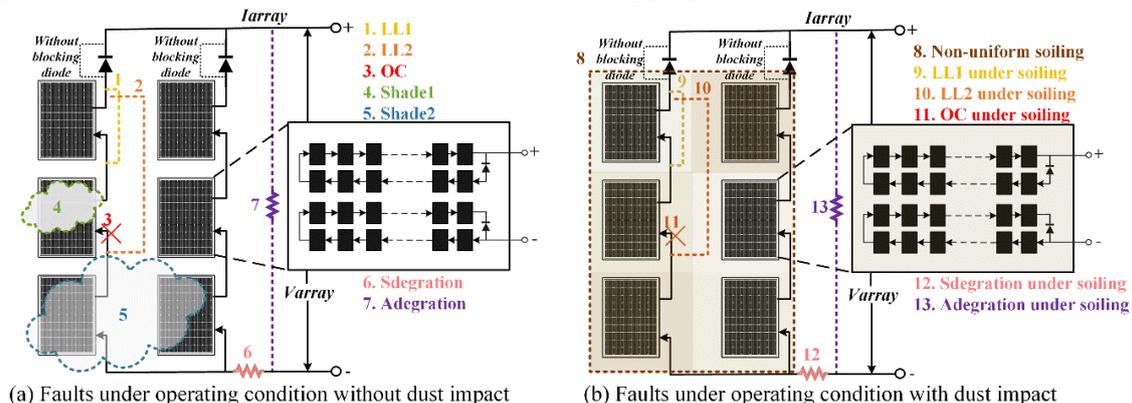

Fig. 1. Faults of PV arrays under different operating conditions.

### 2.1 Faults analysis of PV arrays

As shown in Fig. 1a, two configurations (with and without blocking diodes, denoted as Configuration 1 and Configuration 2, respectively) are designed to analyze each type of fault. In fact, during the operation of a PV array, various faults may occur, including short-circuits in substrings such as short-circuit in one or two modules, open-circuit of substrings, various degrees of shading such as shading of a single module or multiple modules, as well as series resistance degradation of array and parallel resistance degradation of array.

PV arrays with different blocking diode configurations differ in the manifestation of short-circuits [6], as exemplified by the I-V characteristic curves shown in Fig. 2. Specifically, the open-circuit voltage of a normal substring (denoted $V_{OCS}$) of a PV array is proportional to the number of serial modules (denoted $N_{SM}$), and when different numbers of PV modules in a particular substring are short-circuited (denoted $N_{LL}$), then the OC voltage of the fault string is $(N_{SM}-N_{LL})/N_{SM} \cdot V_{OCS}$ (denoted as $V_{OCF}$). When substrings are equipped with blocking diodes, i.e. Configuration 1, the current of each string is only allowed to flow in one direction. $V_{OCS}$ and $V_{MPP}$ of the faulty string are reduced, while $I_{MPP}$ remains unchanged. When $V_{OCS} > V_{OCF}$, the faulty string will be disconnected from array, which is expressed in the IV/PV characteristic curve as the presence of a local minimum inflection point and the overall $V_{OC}$ of the array remaining stable, as shown in Fig. 2b. On the contrary, when the substrings are not equipped with blocking diodes, i.e. Configuration 2, the current from the normal string is reversed into the faulty string when $V_{OCS} > V_{OCF}$. As a result, $V_{OC}$ of the faulty array with short-circuits is lower than that of the normal array. It is worth noting that when $V_{OCS} < V_{OCF}$, the characteristic curves of the PV array are the same for both configurations in this interval.

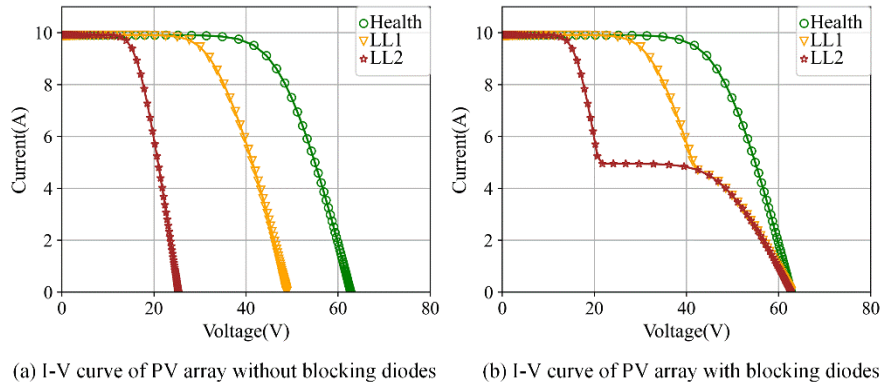

(a) I-V curve of PV array without blocking diodes    (b) I-V curve of PV array with blocking diodes

**Fig. 2.** I-V curve of PV arrays with different blocking diode configurations.

In addition, during outdoor operation, all modules in the PV array can be affected by various types of shading, including shading from buildings, tree shade, etc., as well as shielding from dust adhering to the surface. The essential effect of both is manifested in the reduction of irradiance intensity of the affected modules. Specifically, the former generally results in a larger reduction in irradiance typically of 20% or more, whereas dust accumulation is considered a specific form of shading that occurs across all PV modules of an array and produces a relatively small reduction in irradiance usually below 20% [40, 41].

The enlarged structure of a PV module in Fig. 1a shows that every $N$ cells are inversely connected in parallel with one bypass diode. For a single module, when the shading intensity and area are fixed, the output I-V characteristic curve of the module shows a single peak, as shown in Fig. 3a. Further, when a module in a substring containing multiple modules is shaded, the I-V curve splits into two parts [30]. When the voltage is lower than the substring voltage (denoted $V_S$), the shaded cells of the shaded module are bypassed by the bypass diode, and the I-V characteristic in this interval is approximately equivalent to the I-V characteristic of the remaining unshaded modules in series. When the voltage is larger than $V_S$, the bypass diode turns off resulting in the decrease of the total output current of the string [42]. The I-V curve in this interval is mainly determined by the shaded module. Therefore, no matter what degree of shading, there is a turning peak on the I-V/P-V characteristic curve. The shadow degree can influence the inflection point of current, as shown in Fig. 3a. The current at inflection point decreases with the increase of the shading degree and the number of shading modules. As the degree of shading increases, the voltage at inflection point decreases proportionally, as shown in Fig. 3b. It also can be seen that in Fig. 2 and Fig. 3b, the characteristic curves of arrays under certain shading conditions are very similar to the short-circuited arrays configured with blocking diodes.

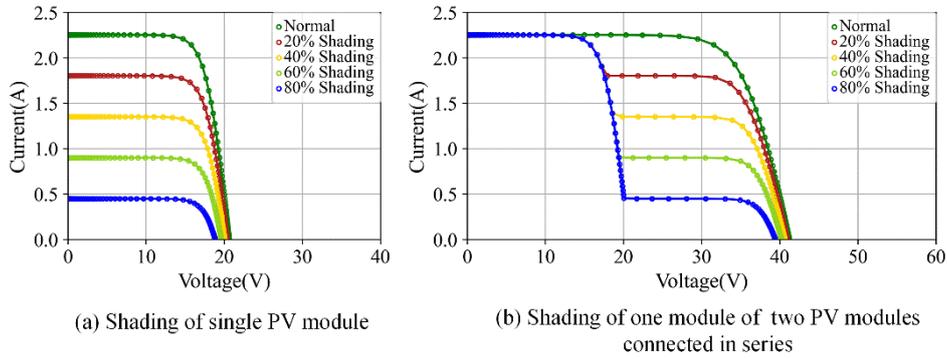

(a) Shading of single PV module

(b) Shading of one module of two PV modules connected in series

**Fig. 3.** I-V curve of PV modules with different structures under the influence of different shading levels.

In fact, the intensity of dust accumulation is uneven across all modules, representing a non-uniform soiling effect for the array [38, 40] as shown in Fig. 1b. Therefore, under the influence of non-uniform soiling, each module of the array is equivalent to being affected by different degrees of shading, the characteristic curves of the array will also show a different number of peaks. Importantly, when other faults occur under the impact of dust, such as short-circuit and open-circuit, the characteristic curves become more complex, showing a superposition of multiple peaks and corresponding fault characteristics at the same time, as in Fig. 4b. However, most cleaning operations for dust or soiling are usually carried out on an annual basis or more frequently in areas heavily affected by dust, but other faults under the influence of non-uniform soiling can still occur during the cleaning interval [43, 44]. This makes the diagnosis of complex fault types more challenging and difficult to identify accurately in time.

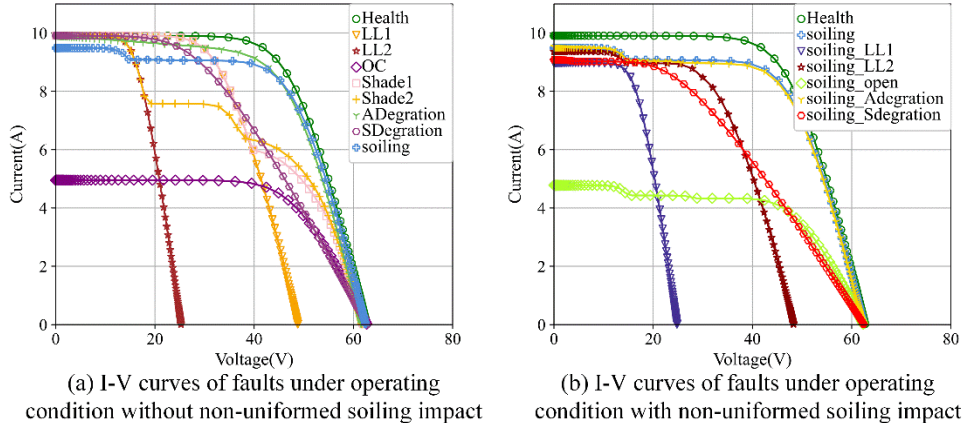

(a) I-V curves of faults under operating condition without non-uniformed soiling impact

(b) I-V curves of faults under operating condition with non-uniformed soiling impact

**Fig. 4.** I-V curves of faults under different operating conditions.

It can be seen that various complex fault types of different array structures are rich and diverse, and have high similarities. Therefore, it is of universal significance to propose an effective method applicable to fault diagnosis in multiple scenarios.

### 2.2 Preprocessing methods for fault diagnosis

The shape of PV characteristic curves is dependent on environmental conditions such as irradiance and temperature [45]. Therefore, when using characteristic curves for PV array fault diagnosis, the influence of different environmental conditions causing different characteristics manifestations of the same fault type should be excluded to reduce confusion of features caused by factors other than the fault types [39]. At present, there are three data preprocessing methods to minimize the impact of environmental factors on the characteristic curves:

(1) The input feature is a two-dimensional matrix recombined by I-V curve and ambient variables, which

stems from the ability of the convolutional neural network to automatically extract features of two-dimensional data. Specifically, irradiance and temperature are repeatedly supplemented into column vectors with the same points of I/V vector, and then stitched together with the I-V curves to form a final two-dimensional matrix in Fig. 5, which is used as input for the deep learning network classification method.

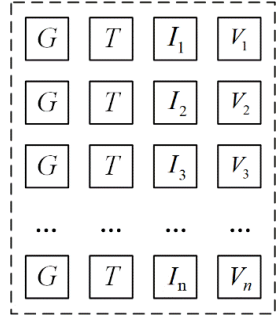

**Fig. 5.** The recombined GTIV matrix.

This method does not directly eliminate the influence of environmental variables on the characteristic curves, but only relies on the convolutional network to extract common feature information of the same fault type under different environmental conditions. It can be inferred that its ability to deal with complicated faults diagnosis is shallow.

(2) Converting key features of array characteristics in different environments to features under STC is often applied to the diagnosis method using key features identified in characteristic curves. That is, the open-circuit voltage $V_{OC}$, the short-circuit current $I_{SC}$, the maximum power point voltage $V_{MPP}$, the maximum power point current $I_{MPP}$, and the equivalent series resistance $R_S$ are converted to corresponding values under STC. The approach to obtain the feature functions is based on the traditional approximation equations [46, 47], where the unknown parameters are denoted by $a$, $b$, $c$ and $d$. The functions of key features can be rewritten as:

$$V_{OC} = V_{OC,STC} + a_1 \cdot \ln\frac{G}{G_{STC}} + a_2 \cdot dT + a_3 \cdot \frac{G}{G_{STC}} dT \tag{1}$$

$$V_m = V_{m,STC} + b_1 \cdot \ln\frac{G}{G_{STC}} + b_2 \cdot dT + b_3 \cdot \frac{G}{G_{STC}} dT \tag{2}$$

$$I_m = c_1 \cdot I_{m,STC} \frac{G}{G_{STC}} + c_2 \cdot dT + c_3 \cdot \frac{G}{G_{STC}} dT \tag{3}$$

$$R_S = R_{S,STC} \left(\frac{G}{G_{STC}}\right)^{d_1} + d_2 \cdot dT + d_3 \cdot \frac{G}{G_{STC}} dT \tag{4}$$

The parameters of this non-linear static model need to be identified by searching through multiple characteristic curves. In addition, as the curves under STC still behave differently for various fault types, separate parameter identifications are required for various faults with different STC conditions, which involves additional test experiments with fault samples. Moreover, the feature-dependent diagnosis method is not effective in classifying complex fault types.

(3) Correcting all points of the entire I-V characteristic curve is a preprocessing method that relies on the use of full I-V information for diagnosis. The IEC 60891 [48] defines three standard procedures for the correction of I-V curves, which are used to allow comparison of curves measured under different conditions, thus enables health monitoring of PV panels. The following are named M1, M2 and M3 respectively, as well as the correction method of M2new proposed in [49]:

M1:

$$I_2 = I_1 + I_{SC} \cdot \left(\frac{G_2}{G_1} - 1\right) + \alpha \cdot T_2 - T_1 \tag{5}$$

$$V_2 = V_1 - R_S \cdot I_2 - I_1 - \kappa \cdot I_2 \cdot T_2 - T_1 + \beta \cdot T_2 - T_1 \tag{6}$$

M2:

$$I_2 = I_1 \cdot [1 + \alpha_{rel} \cdot (T_2 - T_1)] \cdot \frac{G_2}{G_1} \tag{7}$$

$$V_2 = V_1 + V_{OC1} \cdot \left[\beta_{rel} \cdot (T_2 - T_1) + a \cdot \ln\left(\frac{G_2}{G_1}\right)\right] - R_S \cdot (I_2 - I_1) - \kappa \cdot I_2 \cdot (T_2 - T_1) \tag{8}$$

M2new:

It uses the same equation as M2 for current correction, but corrects for voltage by replacing the term "$V_{OC1}$" in (8) with "$V_{OC1} \cdot [1 + \beta_{rel}(25 - T_1)]$".

M3:

$$I_3 = I_1 + \gamma (I_2 - I_1) \tag{9}$$

$$V_3 = V_1 + \gamma (V_2 - V_1) \tag{10}$$

$$G_3 = G_1 + \gamma (G_2 - G_1) \tag{11}$$

$$T_3 = T_1 + \gamma (T_2 - T_1) \tag{12}$$

In fact, M1, M2, and NewM2 are correction methods based on one single curve, that requires the setting of corresponding correction factors such as $\kappa$ and $R_S$. That is, after the correction parameters are determined, they can be directly corrected from any test conditions to the STC. However, it is still difficult to determine the correction factors of PV panels on site due to the rigorous experimental conditions required in the IEC 60891 procedure. In addition, research shows that due to differences in irradiance, module temperature, and severity of faults, all these methods introduce significant errors, making it difficult to perform well under all fault conditions [49]. Moreover, the distortion of curve shape in the IEC method usually leads to a relative error of 13.8% and the estimation error of fault features extracted from the correction curve also occurs frequently. It can be seen that if these features are used as defect features, it may affect the diagnosis of faults.

Alternatively, M3 does not contain correction factors, but the interpolation constant $\gamma$ needs to be set. This method is to apply linear interpolation method on multiple I-V characteristic testing curves to obtain the I-V characteristic under the STC. Although the multiple curves-based method (M3) generally offers higher performance than the above methods based on one single curve, the conversion of a certain curve with both $G$ and $T$ requirements relies on multiple testing curves, and the measurement and calculation are relatively complex and inefficient, so it is not suitable for rapid field diagnosis.

It is undeniable that the use of characteristic curves for PV health monitoring and fault diagnosis is a promising method. Considering the limitations of the practical application of the preprocessing methods discussed above, it will be beneficial to explore solutions based on field measurement data while reducing the dependence on the calibration process.

## 3. The proposed fault diagnosis methodology

In this section, firstly, the accurate modeling approach of the real PV array is presented, and the simulation method of faults in different configurations of PV arrays under different operating conditions is described. Then, a data processing method that does not depend on the calibration of the characteristic curves to STC is proposed to obtain transformed graphical feature matrices with full information of characteristic curves. Finally, the fault diagnosis model based on the convolutional neural network with CBAM module is introduced.

### 3.1 Configuration of the simulated PV modeling

To explore the complex faults of PV arrays under different operating conditions, the single-diode PV cell model proposed in [50] is used in this paper, as shown in Fig. 6. And based on this model a PV module model is built on the PSCAD/EMTDC platform with the configuration parameters from the Shell Solar SP-70 datasheet in Table 1. We compare the simulated I-V curves of this PV module model under different radiation intensities at 25°C with the manual curves supplied by manufacturer. There is a high correlation between the two, which verifies the consistency of the equivalence between the model and the actual PV module. Based on this PV module model, the PV array model in this study is further designed for a structure with three modules in series and two substrings in parallel. Among them, the substrings containing blocking diodes or not results in the two configuration types of PV arrays, whose structure is shown in Fig. 1. In order to characterize the faults under the operating environmental conditions of the real PV array as much as possible, we take one year of outdoor irradiance and temperature records measured from the PV plant as the environmental control variables for the PV array model.

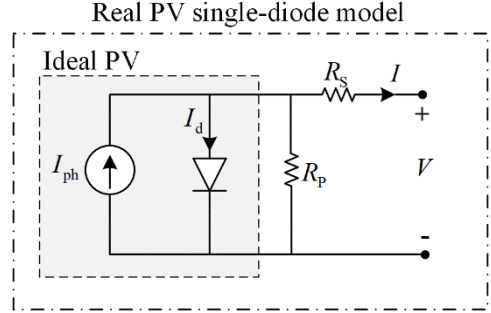

**Fig. 6.** The single-diode PV cell model.

**Table 1** Electrical characteristics of solar module Shell SP-70 at nominal condition (25°C and 1000 W/m$^2$).

| Parameter | Value |
| --- | --- |
| $I_{SC}$ (A) | 4.7 |
| $V_{OC}$ (V) | 21.4 |
| $I_{MPP}$ (A) | 4.25 |
| $V_{MPP}$ (V) | 16.5 |
| $K_V$ (mV/°C) | -76 |
| $K_i$ (mA/°C) | 2 |
| $N_S$ | 36 |
| $R_S$ (Ω) | 0.41 |
| $R_P$ (Ω) | 141 |

To fully demonstrate the generalizability of the proposed fault diagnosis method, the faults of PV arrays with two types of blocking diode configurations under different operating conditions are validated, including contamination-prone operating condition and ideal normal operating condition. The two operating conditions consist of 14 and 9 types of faults, respectively, as follows.

**Case1: PV arrays under operating condition with non-uniform soiling impact**

1) Two types of line-to-line short-circuit (LL): one or two modules in one string are shorted (noted as LL1 and LL2 respectively).

2) Open-circuit (OC): one string is open.

3) Two types of shading (Shade): 1 or 2 modules are shaded to different degrees (noted as Shade1 and Shade2, respectively).

4) Series resistance degradation of the array (Sdegration): an increase in the equivalent series resistance of array.

5) Parallel resistance degradation of the array (Adegration): a decrease in the equivalent parallel resistance of array.

6) Non-uniform soiling (Soiling): the accumulation of soiling with varying degrees on the surface of each PV module in the presence of contamination.

7) Line-to-line short-circuit under the impact of non-uniform soiling (soiling_LL): 1 or 2 modules in one string are shorted under varying soiling accumulation (denoted as soiling_LL1 and soiling_LL2, respectively).

8) Open-circuit under the impact of non-uniform soiling (soiling_OC): one string is open under varying soiling accumulation.

9) Series resistance degradation of the array under the impact of non-uniform soiling (soiling_Sdegartion): an increase in the equivalent series resistance of array under varying soiling accumulation.

10) Parallel resistance degradation of the array under the impact of non-uniform soiling (soiling_Adegartion): a decrease in the equivalent series resistance of array under varying soiling accumulation.

**Case2：PV arrays under operating condition without non-uniform soiling impact**

The types of faults include 1) - 6) above under the ideal operating condition without the influence of non-uniform soiling.

As an example, the characteristic curves for faults of the PV array with the two blocking diode configurations in Case1, including I-V and P-V are shown in Fig. 7 and Fig. 8.

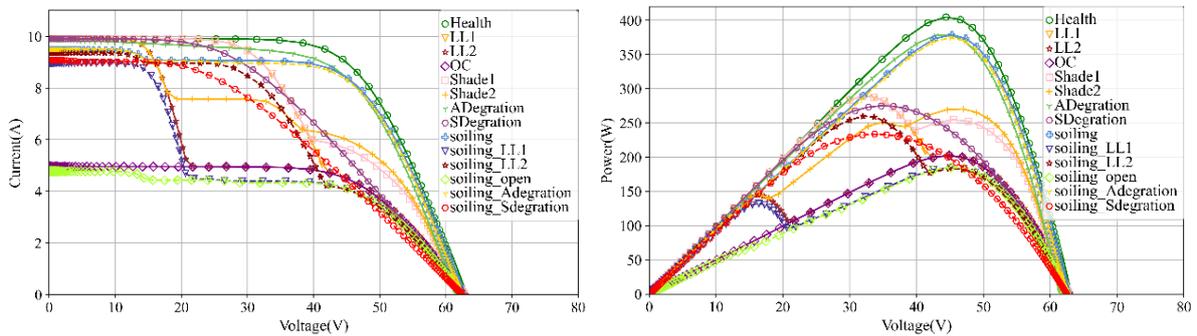

**Fig. 7.** Faults characteristic curves of PV array with blocking diodes.

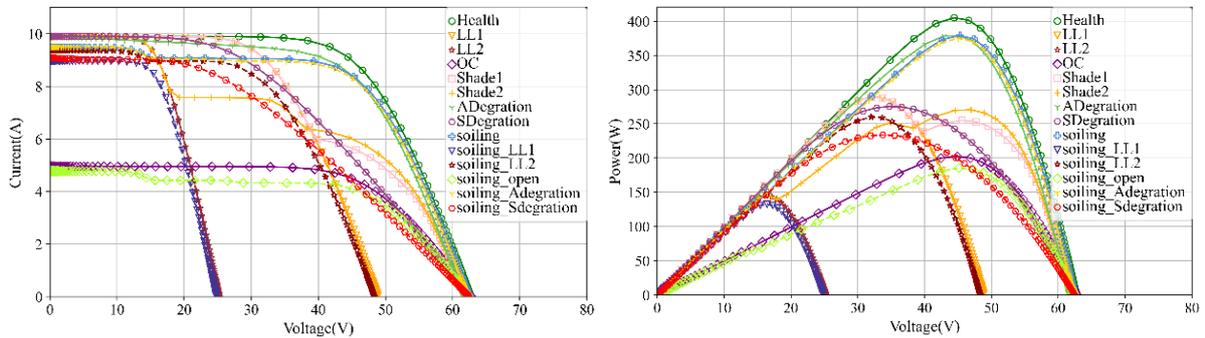

**Fig. 8.** Faults characteristic curves of PV array without blocking diodes.

Importantly, what is different from [18, 21, 37] is that the simulations of faults in this study for Shade, ADegration, SDegration, and Soiling, as well as the severity of these faults, are not simply set as constants. This is due to the fact that these faults change dynamically with time in service, therefore the simulations should contain the full range of fault parameters. In this study, the environmental conditions of a real PV plant are used as inputs for the temperature and irradiance of the PV array model, and the fault parameter of each fault type is set randomly within its corresponding full-scale parameter range to obtain characteristic curves. Specifically, for shading, the irradiance gain is set from 20% (low shading) to 100% (full shading) for one or two modules; for soiling, a special form of shading, the irradiance gain is set randomly within 10% for all modules to simulate non-uniform soiling accumulation; for abnormal aging of ADegration and SDegration, the aging resistance values are set randomly for degradation in [20Ω, 200Ω] and [1Ω, 15Ω] respectively, adjusting for different levels of fault severity. Considering the dust-influenced operating condition, faults are simulated by superimposing other faults alongside the non-uniform soling, where the degree of soiling and dynamic fault parameters are set randomly when it comes to dynamically growing fault types. The representation of different faults with full-scale dynamic fault parameter is shown in Fig. 9, where the characteristic curves show different shapes of distortions.

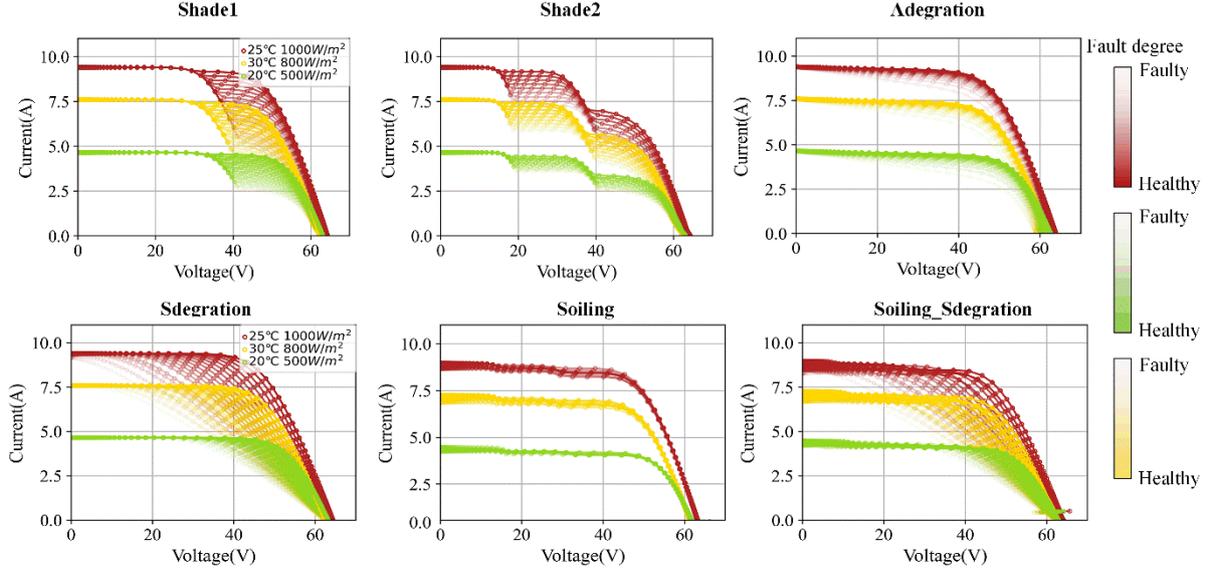

**Fig. 9.** I-V curves of different faults with full-scale dynamic fault parameter.

### 3.2 Graphical feature transformation method of Isc-Voc normalized GADF

The graphical feature transformation method using the Gramian angular difference field (GADF) [51] enables the extraction of complete information from the resampled characteristic curves. Typically, GADF is applied to the transformation of time series signals, which mainly includes the following steps:

Step 1: Scale the data to [0, 1] according to the following equation.

$$\tilde{x}_i = \frac{x_i - \min(X)}{\max(X) - \min(X)} \tag{13}$$

where, the time series data is $X = \{x_1, x_2, ..., x_N\}$ and its value at each timestamp is $x_i$, $\tilde{x}_i$ is noted as the normalized value.

Step 2: Convert the rescaled sequence to the polar coordinate system, i.e. the values of the time series are treated as the cosine of the angle. The formulas for converting to polar coordinates are:

$$\begin{cases} \phi_i = \arccos \tilde{x}_i, 0 \leq x_i \leq 1, \tilde{x}_i \in \tilde{X} \\ r_i = \frac{t_i}{N}, t_i \in N \end{cases} \tag{14}$$

where, $t_i$ is the time stamp of point $x_i$, $N$ is the number of all the time points contained in the time series, and $\tilde{X}$ represents the rescaled time series.

Step 3: Obtain the angle difference of each pair, and then take the sine value for difference to form the GADF matrix:

$$GADF = \left[\sin\left(\phi_i - \phi_j\right)\right] = \sqrt{I - \tilde{X}^2} \cdot \tilde{X} - \tilde{X}' \cdot \sqrt{I - \tilde{X}^2} \tag{15}$$

where, $I$ is the unit row vector [1, 1, ..., 1], $r_i$ is the value of the polar axis, and $\phi_i$ the value of the polar angle.

It is worth noting that when applied to time series, the scaling of GADF is using the full-time axis data, and thus when applied to the scaling of the I-V/P-V characteristic curves, the V-axis should be analogous to the time axis. If GADF with normal normalization following steps described above is used to convert the characteristic curves, i.e. the conversion areas are A, B, C and D respectively in Fig. 10, in essence, the current $V_{OC}$ and $I_{SC}$ of each fault state are taken as their respective normalized maximum values, which would blur the differences in $V_{OC}$ and $I_{SC}$ between the different faults. We can observe from the GADF transformation results in Fig. 11 that when GADF transformation is performed on the I-axis or P-axis with the normal normalization strategy, the same fault types under different irradiance and temperature conditions are relatively consistent in the shape of the converted graphical features, which can avoid misidentification of the same faults due to different environments. However, similarities in shapes also exist between the characteristic curves of certain different fault types, such as health, LL1, LL2, and OC in Fig. 11. In fact, these curves differ in absolute values of $V_{OC}$ and $I_{SC}$. Therefore, it is necessary to propose a new GADF transformation method that retains consistency in the characteristics of the transformation features for the same fault in different environmental conditions, while having the ability to distinguish the similarity in the shapes of the characteristic curves for different fault types.

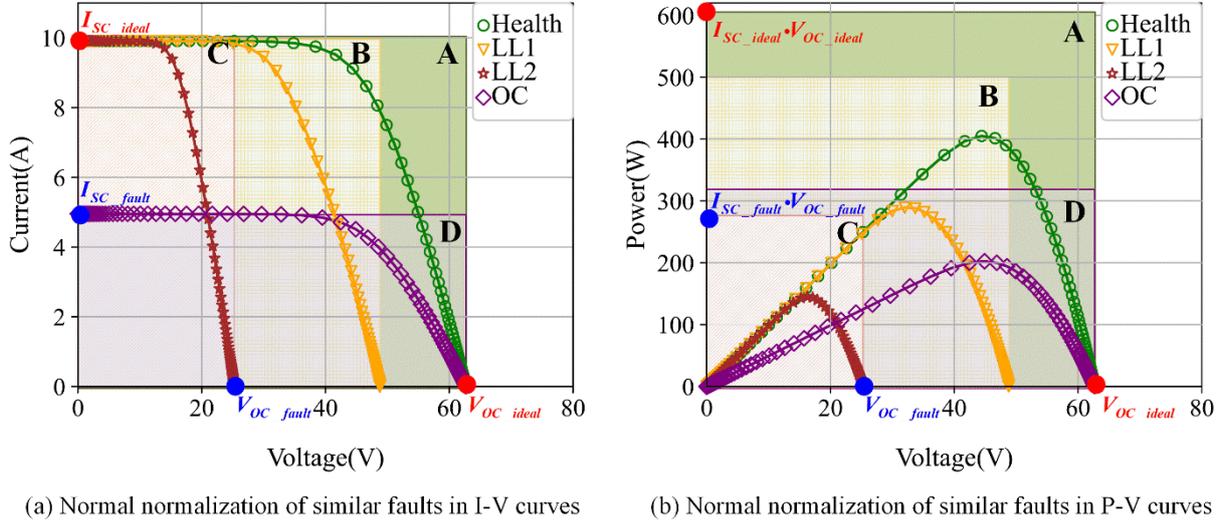

(a) Normal normalization of similar faults in I-V curves  (b) Normal normalization of similar faults in P-V curves

**Fig. 10.** Normal normalization area of similar faults under the same ambient condition.

Through the analysis of the single-diode equivalent model of the PV module in Fig. 6, the calculation equations for the short-circuit current $I_{SC}$ and the open-circuit voltage $V_{OC}$ are obtained as follows:

When the external circuit is short-circuited, i.e. the load is 0, the short-circuit current $I_{SC}$ is:

$$I_{SC} = I_{ph} - I_0 \left[ e^{\frac{I_{SC}R_S}{nVT}} - 1 \right] - \frac{I_{SC}R_S}{R_P} \tag{16}$$

The dark current $I_d = I_0 \left[ e^{\frac{I_{SC}R_S}{nVT}} - 1 \right]$ flowing through the diode is very small and can be ignored. Considering that $R_P \gg R_S$, $\dfrac{I_{SC}R_S}{R_P}$ can also be ignored [47], then:

$$I_{SC} \approx I_{ph} \tag{17}$$

$$I_{ph} = \left[ I_{SC,n} + k_i \cdot (T - 298) \right] \cdot \frac{G}{1000} \tag{18}$$

where, $T$ is the temperature in Kelvin, $G$ is the irradiance, and $k_i$ is the temperature coefficient of short-circuit current.

When the external circuit is open, i.e. when the load is close to $\infty$ and $I = 0$, the open-circuit voltage $V_{OC}$ is [52]:

$$I = I_{ph} - I_0 \left[ e^{\frac{I_{SC}R_S}{nVT}} - 1 \right] - \frac{I_{SC}R_S}{R_P} \tag{19}$$

$$0 = I_{ph} - I_0 \left[ e^{\frac{V_{OC}}{nVT}} - 1 \right] - \frac{V_{OC}}{R_P} \tag{20}$$

$$V_{OC} = \frac{nkT}{q} \ln \left( \frac{I_{ph}}{I_0} + 1 \right) \tag{21}$$

where, the diode saturation current $I_0$ is expressed as [53, 54]:

$$I_0 = I_{0,n} \left(\frac{T_n}{T}\right)^3 \exp\left[\frac{qE_g}{nk}\left(\frac{1}{T_n} - \frac{1}{T}\right)\right] \tag{22}$$

$$I_{0,n} = \frac{I_{SC,n}}{\exp V_{OC,n}/nV_{t,n} - 1} \tag{23}$$

$$V_{t,n} = \frac{N_S k T_n}{q} \tag{24}$$

where the band gap energy of the semiconductor is represented by $E_g$ and the nominal saturation current at the nominal temperature $T_n$, i.e. 25°C, is represented by $I_{0,n}$ via (23), the nominal thermal voltage represented by $V_{t,n}$ is shown in (24), $I_{SC,n}$ is the short-circuit current at nominal temperature, and $V_{OC,n}$ is the open-circuit voltage at nominal temperature. In addition, $n$ is the ideal factor of the diode, $N_S$ is the number of series cells forming the module, $k$ is the Boltzmann constant and $q$ is the charge quality. Thus, it can be seen from (22) that $I_0$ is only related to the temperature $T$.

It can be further concluded from (18) and (21) that the short-circuit current $I_{SC}$ and open-circuit voltage $V_{OC}$ of PV modules depend only on irradiance $G$ and ambient temperature $T$, except for the constant factors with fixed values, which means that the ideal $I_{SC}$ and $V_{OC}$ are the same for different fault types under the same environmental conditions. Therefore, we propose a GADF method based on the normalization of $I_{SC}$ and $V_{OC}$ under environmental conditions, which can ensure that the transformed graphical features of the same fault types are the same in different environments, while having distinguishability between different faults with similar forms. Specifically, the main implementation process of the Isc-Voc normalized GADF method is:

(1) Calculate the ideal $I_{SC}$ and $V_{OC}$ (denoted as $I_{SC\_ideal}$ and $V_{OC\_ideal}$) of the ambient conditions corresponding to each characteristic curve, expand the maximum value of V-axis to the ideal $V_{OC}$, and complement the current and power of the characteristic curve by 0 over the range of the measured $V_{OC}$ and the ideal $V_{OC}$.

(2) Resample the complemented characteristic curves using the bilinear difference method to provide uniformly distributed characteristic curves with a small amount of data. Specifically, the V-axis of I-V and P-V curves is within the range of [0, $V_{OC\_ideal}$] at a uniform voltage interval and the I-axis and P-axis are resampled at the same uniform interval, reducing data of each curve from the original 200 points to 50 points.

(3) Convert the resampled characteristic curves to graphical features according to Isc-Voc normalized GADF. That is, the V-axis is normalized according to the range of [0, $V_{OC\_ideal}$], the I-axis is normalized according to the range of [0, $I_{SC\_ideal}$], and the P-axis is normalized according to the range of [0, $V_{OC\_ideal}$ * $I_{SC\_ideal}$].

(4) Then, calculate the inner product according to the differences of normalized angles, preserving the time dependence of the V-axis, and generate a GADF matrix with size of $50^2$. The I-V or P-V characteristic curve corresponding to each environmental condition is transformed into a matrix, and the I-V and P-V transformation matrices are stacked to form a 2-channel 2D matrix as the input to the fault diagnosis model.

This method makes it possible to select a unified $V_{OC}$ under the same environmental conditions, which corresponds to the V-axis having the same time scale. In this way, all changes in characteristic curves can be reflected in the transformation matrices. When the slope of the characteristic curve changes, the diagonal regions in the matrix shrink in different directions. Further, the normalization strategy of selecting the maximum value of $V_{OC}$ and the maximum value of $I_{SC}$ in all characteristic curves, without distinguishing environmental conditions, is also compared and referred to as the global normalized GADF. More detailed discussion and analysis of the three current universal applied methods (direct I-V, RP, and GADF) as well as the proposed normalization strategies are provided in Chapter 4.

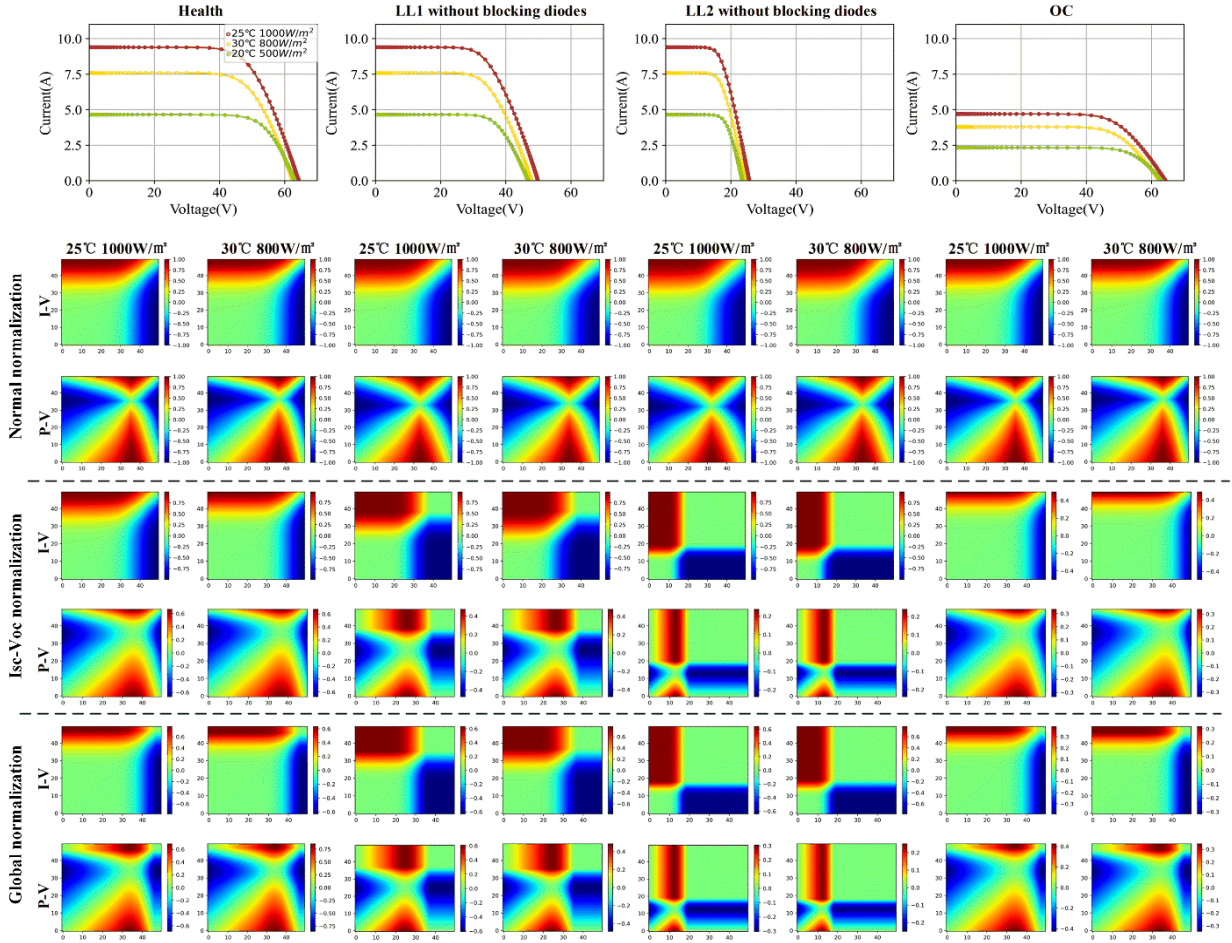

**Fig. 11.** Graphical matrices transformed by GADF with different normalization strategies for similar faults under different environmental conditions (the matrices are colored for only visualization).

### 3.3 Classification model of convolutional neural network with CBAM module

To improve the ability of diagnosing complicated faults, we design a PV array fault diagnosis model of convolutional neural network with CBAM module, referred to as CNN-CBAM, with the transformed graphical feature matrices containing full information of the characteristic curves as the input features.

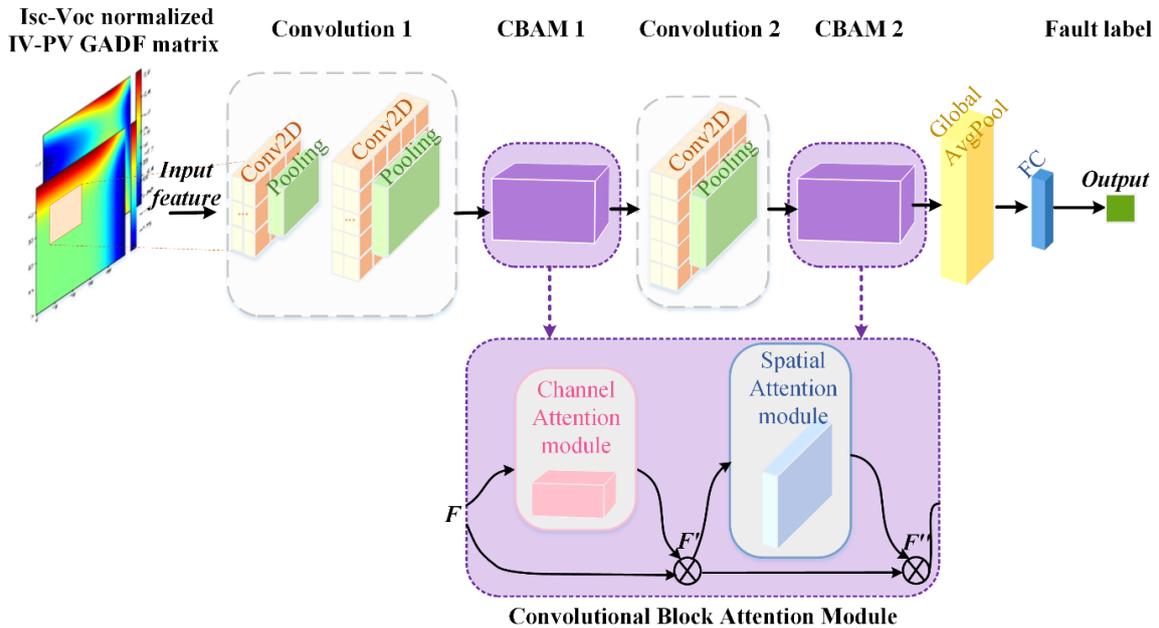

**Fig. 12.** The structure of proposed classification model of convolutional neural network with CBAM module.

The structure of the model is shown in Fig. 12, which is mainly composed of convolution modules and CBAM attention modules.

The Convolution module can reduce the dimensions in time and space and lower the number of free parameters required for training [55], due to the benefits of local receptive field and weight sharing. Therefore, performance can be enhanced. Specifically, the Conv2d layer slides each filter in the input feature matrix through the local receptive field, calculates the sum of dot products on the local field, and automatically extracts the effective features from the inputs. Then, the Pooling layer divides the input area and calculates the average value of each area to complete the down sampling of the feature matrix.

The CBAM module focuses on the important features and suppresses the unimportant features in the network, which can effectively improve the performance of CNN. And the CBAM is composed of the channel attention module (CAM) and the spatial attention module (SAM) [56]. Their detailed structures are shown in Fig. 13 and Fig. 14, respectively. Among them, the CAM emphasizes that the network should concentrate on useful channel features while ignoring other aspects by using the maximum pool and the average pool to compress the spatial dimension of the feature matrix. The SAM highlights that the network should focus on the local area of interest by applying the average pool and maximum pool along the channel dimension to retain the background information of the feature matrix. Specifically, first, the input feature $F$ is multiplied by the feature matrix $M_C(F)$ generated by CAM compression along the spatial dimension to obtain $F'$. Then, SAM compresses $F'$ along the channel dimension to generate the spatial feature matrix $M_S(F')$. Finally, the optimized feature matrix $F''$ is obtained by multiplying $M_S(F')$ and $F'$. The whole process can be represented by:

$$F' = M_C(F) \otimes F \tag{25}$$

$$F'' = M_S(F') \otimes F' \tag{26}$$

where, $\otimes$ represents multiplication between elements

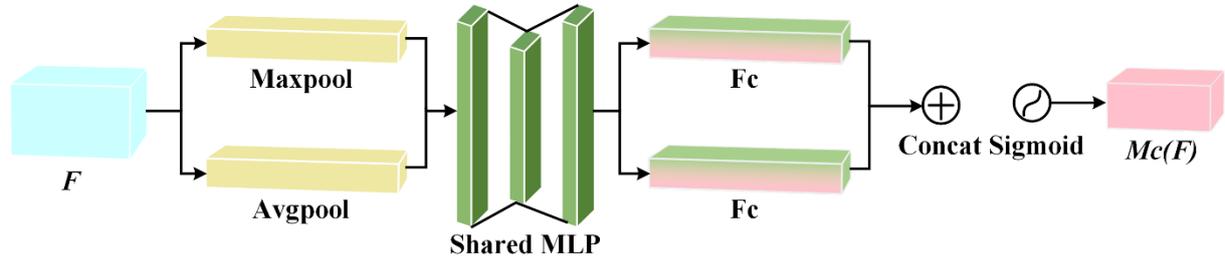

Fig. 13. The structure of channel attention module (CAM).

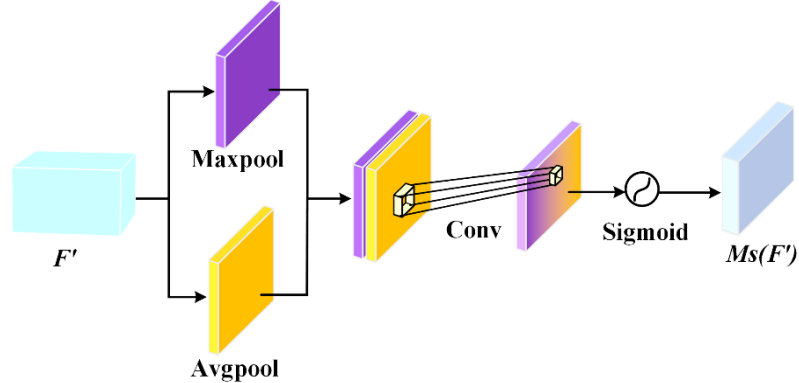

Fig. 14. The structure of spatial attention module (SAM).

The CAM module processes and aggregates the input features with the maximum pool and average pool, and then inputs them into a weight-sharing multi-layer perceptron (MLP) network for summation, which is activated by sigmoid to generate the final channel attention feature $M_C(F)$. The calculation program of CAM can be expressed as:

$$M_C(F) = \sigma(MLP(AvgPool(F)) + MLP(MaxPool(F))) \tag{27}$$

The SAM module performs maximum pool and average pool operations on the input feature $F'$ along the channel dimension, generates a two-layer feature matrix, and cascades them together. Then, the convolution kernel with size of 7×7 is used to reduce the dimension of features, and the sigmoid is applied to generate the spatial attention feature $M_S(F')$. The calculation program of SAM can be expressed as:

$$M_S(F')=\sigma(f([AvgPool(F');MaxPool(F')])) \qquad (28)$$

where, σ denotes the sigmoid function and $f$ denotes the convolution operation with filter.

Table 2 shows the specific structural configuration of the proposed PV array fault diagnosis model based on CNN and CBAM. The Convolution blocks with different sizes of convolutional kernels are stacked together along the depth direction to extract features at different levels. This structure can automatically extract more effective feature information directly from the original input feature matrix. In this study, two CBAM modules are respectively embedded into the convolution modules, and the classification accuracy of the faults diagnosis model is improved through the processing capability of focused feature information.

Table 2 Detailed configuration of the CNN-CBAM.

| Layer | Output shape | Detailed structure |
|---|---|---|
| Input Layer | (50, 50, 2) | |
| Convolution 1 | (46, 46, 32) | $k = 3×3$, $filter = 8$, $stride = 1×1$, $padding = 1$ $k = 3×3$, $filter = 32$, $stride = 1×1$, $padding = 1$ |
| CBAM 1 | (46, 46, 32) | |
| Convolution 2 | (44, 44, 64) | $k = 3×3$, $filter = 64$, $stride = 1×1$, $padding = 1$ |
| CBAM 2 | (44, 44, 64) | |
| Global Avgpool | 64 | |
| Fully connected | 16 | |
| Output Layer | classes | |

## 4. Results and discussion

### 4.1 Experimental setup

The data used in this study are based on the configurations described in Section 3.1. Specifically, the faults of PV arrays with two blocking diode configurations under various operating conditions are analyzed, including 14 faults under contamination-prone operating condition and 9 faults under ideal operating condition. To make the data fully reflect the real operating conditions, we take the collected annual ambient records of the actual power station as the environmental control input for each fault type to obtain the corresponding characteristic curves. The data is divided into training set and testing set, accounting for 80% and 20% respectively. And 90% of the training set is the training data and 10% is the validation data. The data volume and proportion of each data set applied to different operating conditions are shown in Table 3.

Table 3 The specific information of the different dataset.

| | | PV array with blocking diodes | | | PV array without blocking diodes | | |
|---|---|---|---|---|---|---|---|
| | | 80% | | 20% | 80% | | 20% |
| | | Training data (90%) | Validation data (10%) | Testing data | Training data (90%) | Validation data (10%) | Testing data |
| Considering soiling impact | Number of classes | 14 | 14 | 14 | 14 | 14 | 14 |
| | Number of data | 43192 | 4800 | 11998 | 43192 | 4800 | 11998 |
| Without considering soiling impact | Number of classes | 9 | 9 | 9 | 9 | 9 | 9 |
| | Number of data | 27766 | 3086 | 7713 | 27766 | 3086 | 7713 |

### 4.2 Faults diagnosis evaluation indexes

The widely used Precision, Recall, F1-score, and Accuracy are selected as the indicators of the effectiveness of the classification algorithm [57, 58], which can be calculated as:

$$\text{Precision} = \frac{TP}{TP + FP} \qquad (29)$$

$$\text{Recall} = \frac{TP}{TP + FN} \quad (30)$$

$$\text{F1\_score} = \frac{2 \cdot \text{Recall} \cdot \text{Precision}}{\text{Recall} + \text{Precision}} \quad (31)$$

$$\text{Accuracy} = \frac{TP + TN}{TP + FP + FN + TN} \quad (32)$$

where, the number of samples that belong to the positive category and are predicted to be in the positive category are referred to as true positives (*TP*), the number of samples that belong to the positive category and are predicted to be in the positive category are referred to as false negatives (*FN*), the number of samples that belong to the negative category and are predicted to be in the positive category are referred to as false positives (*FP*), and the number of samples that belong to the negative category and are predicted to be in the negative category are referred to as true negatives (*TN*).

**4.3 Case 1. PV arrays under operating condition with non-uniform soiling impact**

This part analyzes the effectiveness of different methods to distinguish the 14 fault types of PV arrays under Case 1 operating condition with non-uniform soiling impact, including the accuracy and performance of different graphical feature transformation methods and classification algorithms for fault diagnosis.

**4.3.1 Analysis of graphical feature transformation methods**

First, we compare data preprocessing methods for fault diagnosis that do not rely on STC correction, as presented in Section 2.2, including methods that combine I-V curves and environmental variables to form a two-dimensional data matrix (GTIV) and methods that utilize the complete feature information of the characteristic curves. In addition, the diagnosis results of different graphical feature transformation methods such as GADF and recurrent plot (RP) using the complete information of the characteristic curves are analyzed, and three strategies of using the I-V graphical feature matrix, P-V graphical feature matrix and IV-PV stacked graphical feature matrix obtained from transformation as input are further compared.

Table 4 The fault diagnosis results of applying different data preprocessing methods.

| | | | Precision | Recall | F1-score | Accuracy |
|---|---|---|---|---|---|---|
| Circumstance 1: With blocking diodes | GADF | I-V | 98.40% | 98.31% | 98.35% | 98.46% |
| | | P-V | 95.27% | 95.16% | 95.21% | 95.33% |
| | | IV-PV | **98.53%** | **98.48%** | **98.50%** | **98.58%** |
| | Recurrent Plot | I-V | 80.97% | 80.03% | 80.50% | 81.77% |
| | | P-V | 86.87% | 86.27% | 86.57% | 86.98% |
| | | IV-PV | 94.43% | 94.23% | 94.33% | 95.12% |
| | Direct IV | GTIV | 95.86% | 95.79% | 95.82% | 95.98% |
| Circumstance 2: Without blocking diodes | GADF | I-V | **98.38%** | **98.25%** | 98.31% | **98.50%** |
| | | P-V | 91.56% | 91.32% | 91.44% | 92.06% |
| | | IV-PV | **98.38%** | **98.28%** | **98.33%** | 98.48% |
| | Recurrent Plot | I-V | 67.04% | 65.53% | 66.28% | 65.94% |
| | | P-V | 79.32% | 78.25% | 78.78% | 78.69% |
| | | IV-PV | 90.08% | 89.86% | 89.97% | 90.85% |
| | Direct IV | GTIV | 95.61% | 95.52% | 95.56% | 95.65% |

Table 4 shows the results of applying different data preprocessing methods to the characteristic curves for fault diagnosis, all of which use the multilayer-CNN classifier. It can be seen that when the graphical feature matrix transformed by single curve is used as input, i.e. the I-V or the P-V characteristic curve, the effectiveness of the information contained in the transformation matrix in distinguishing fault types is related to the feature transformation methods. Sp ecifically, the I-V graphical feature of the GADF transformation is more effective in classification than the P-V graphical feature, whereas the opposite is true for the RP transformation. In fact, the classification accuracy of the IV-PV stacked graphical feature matrix is usually higher than that of the single characteristic curve transformed matrix. Additionally, for the faults classification of PV array considering the impact of dust, the accuracies of the graphical feature matrix transformed by GADF are 98.58% and 98.48% respectively in the configurations of with and without blocking diodes, which is better than the RP transformation and GTIV matrix methods. Fig. 15 shows the classification accuracy of the optimal feature strategy for the three preprocessing methods, where the stacked IV-PV matrix is used for graphical feature transformation method. Among them, the accuracy of RP transformation method for arrays with blocking diodes is 95.12%, which is significantly higher than 90.85% without blocking diodes. This may be related to the similarity between the two types of short-circuit and the health state without blocking diode configuration when RP transformation is applied, while the GADF transformation is applicable to different PV array configurations, with high accuracy of 98.58% and 98.48% respectively.

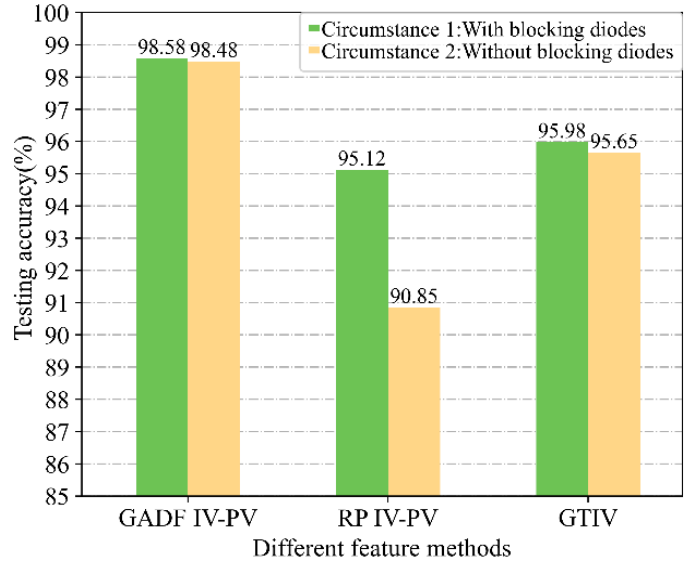

**Fig. 15.** Comparison of the optimal feature strategies for three preprocessing methods.

The above comparative analysis confirms the effectiveness of the GADF graphical feature transformation method for fault diagnosis of PV arrays with different blocking diode configurations. Further, Table 5 compares different GADF normalization approaches. The application of GADF with good diagnostic accuracy first presented in [39] was based on the ANN method, while more types of faults and higher complexity of faults are involved in this study, therefore the classification effects of multilayer-CNN and ANN are compared.

**Table 5** The fault diagnosis results of applying different GADF normalization methods.

|  |  | GADF | Testing accuracy | | |
|---|---|---|---|---|---|
|  |  |  | Normal normalization | Global normalization | Isc-Voc normalization |
| Circumstance 1: With blocking diodes | ANN | IV | 84.52% | 86.81% | 95.66% |
|  |  | PV | 80.44% | 80.36% | 92.97% |
|  |  | IV-PV | 91.42% | 87.40% | **96.11%** |
|  | Multilayer CNN | IV | 98.46% | 98.25% | 98.85% |
|  |  | PV | 95.33% | 91.54% | 97.83% |
|  |  | IV-PV | 98.58% | 98.19% | **99.10%** |
| Circumstance 2: Without blocking diodes | ANN | IV | 72.30% | 83.66% | 94.97% |
|  |  | PV | 65.54% | 75.40% | 91.58% |
|  |  | IV-PV | 87.92% | 82.96% | **95.22%** |
|  | Multilayer CNN | IV | 98.50% | 98.04% | 98.42% |
|  |  | PV | 92.06% | 91.23% | 97.25% |
|  |  | IV-PV | 98.48% | 97.79% | **98.65%** |

The results in Fig. 16 show that for both blocking diode configurations, the graphical feature transformation preprocessing method using Isc-Voc normalized GADF is significantly more accurate in classifying faults than either the normal normalization or global normalization strategies. In particular, when classifying with relatively simple methods such as ANN, the classification effect mainly depends on the effectiveness of the features in distinguishing faults. In the case of array with blocking diodes, for example, the classification accuracy of the Isc-Voc normalized GADF transformation is as high as 96.11%, while the highest classification accuracy of other normalization strategies is only 81.42%. This also fully demonstrates that the Isc-Voc normalization strategy can obtain features that make the GADF transformation matrix of different fault types better distinguishable compared to other normalization strategies. Moreover, the classification accuracy of the CNN is significantly improved over ANN when using the feature matrix of I-V or P-V obtained by normal normalization or global normalization strategies as input, with the diagnostic accuracy of the array without blocking diodes improved by 10.56% and 14.83% respectively. Further, when the I-V/P-V matrix or stacked IV-PV matrix processed by Isc-Voc normalized GADF is used as the input features for the CNN model, they both have higher diagnostic accuracy, which is related to the stronger learning capability of the multilayer CNN.

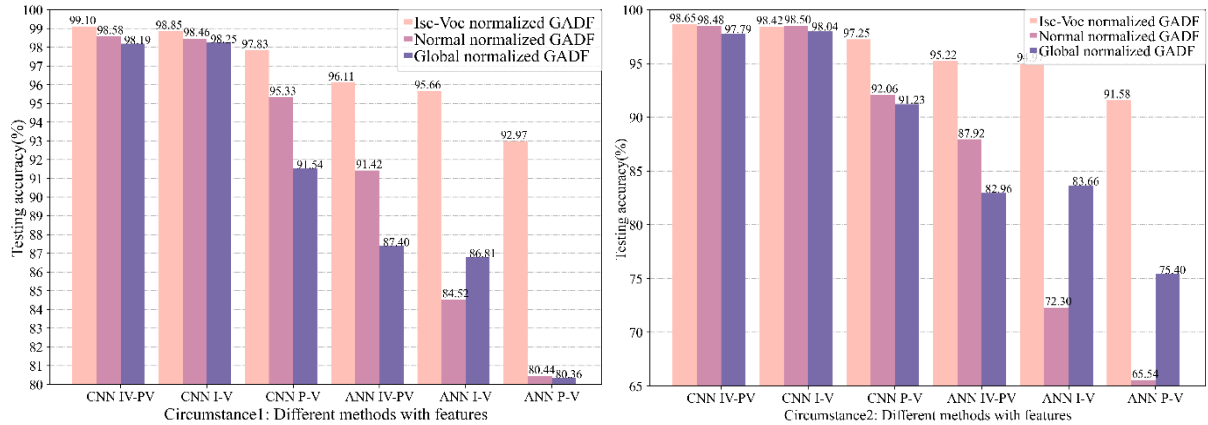

**Fig. 16.** Comparison of three GADF normalization methods with different input matrices and classifiers.

The loss and accuracy of training and validation in Fig. 17 show that using the IV-PV stacked feature matrix can quickly achieve higher classification accuracy, indicating that the stacked matrix has the advantage of more significant differentiation than the others. Similarly, when the features processed by the Isc-Voc normalization strategy are used as the input of the classification model, the training accuracy of the model in Fig. 18 is rapidly improved, and it takes less time to train to the highest accuracy than the features processed by other normalization strategies. In other words, the proposed Isc-Voc normalized GADF graphical feature matrix is more efficient and more accurate when applied to fault diagnosis.

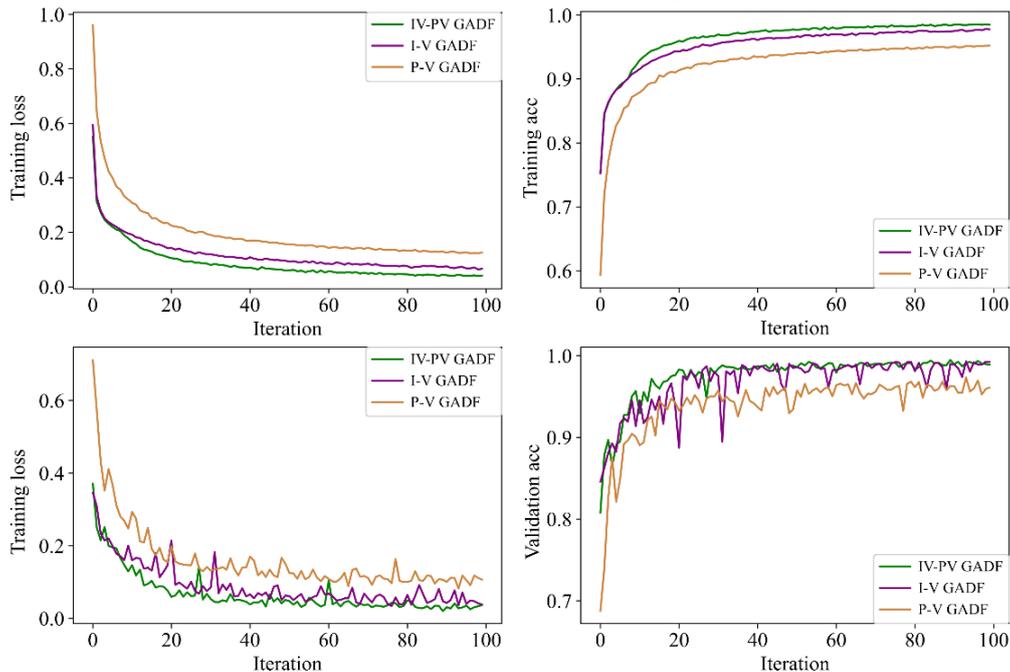

**Fig. 17.** The loss and accuracy for three input matrices.

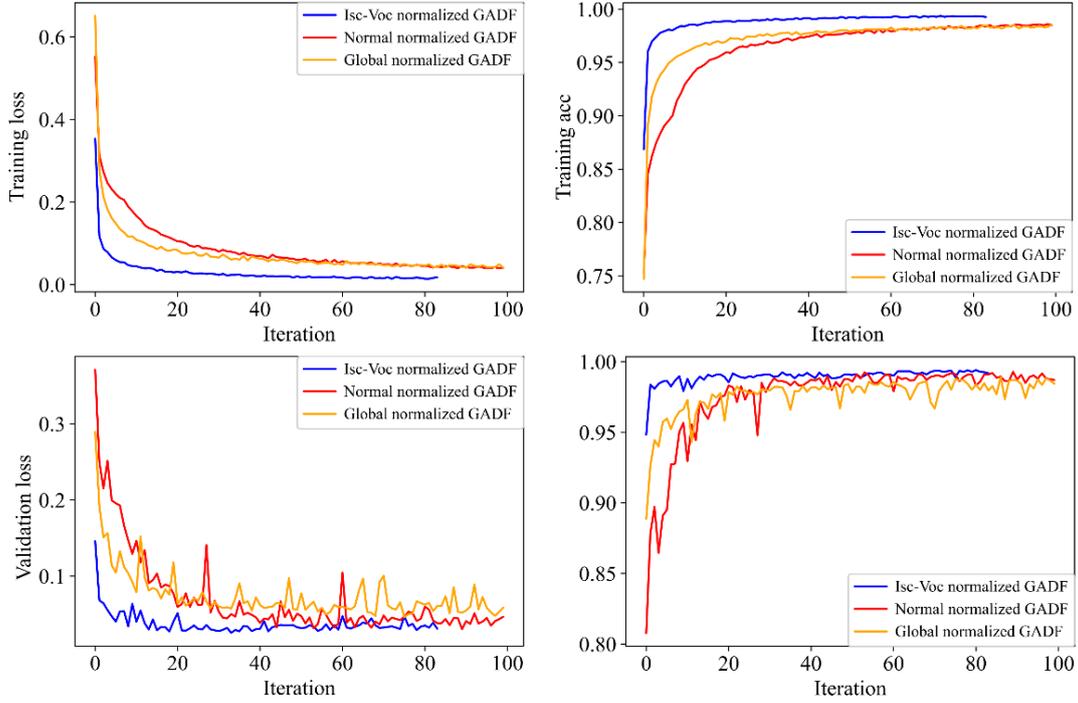

**Fig. 18.** The loss and accuracy for three GADF normalization methods.

**4.3.2 Analysis of convolutional neural network classification models**

Although the classification accuracy of multilayer CNN in the previous section is relatively considerable, the structure of this model is simple and there is still room for improvement. In this section, we further explore the improvement of classification accuracy of CNN-based models with different structures. Therefore, we design and compare the basic multilayer convolutional neural network, the multi-scale convolutional neural network presented in [37], and the proposed convolutional neural network with CBAM module, whose structures are shown in Fig. 19.

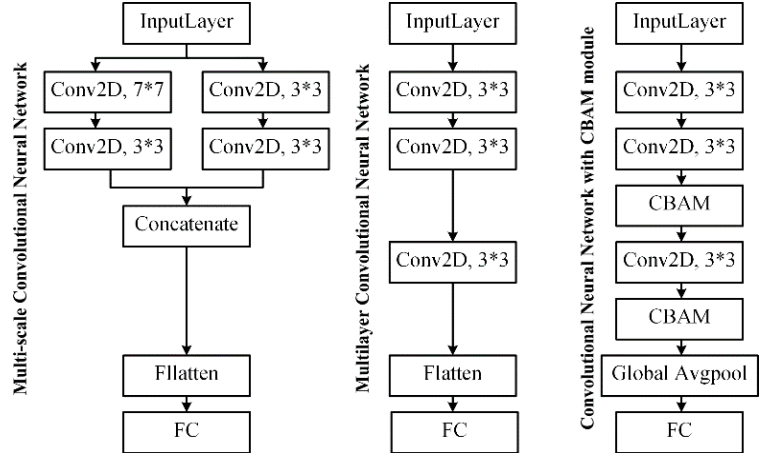

**Fig. 19.** The structures of compared three CNN-based classification models.

The results from Table 6 show that the IV-PV stacked feature matrix obtained from the Isc-Voc normalized GADF transformation has the highest classification accuracy as input to the CNN-based classification models for all three structures, regardless of the blocking diode configuration of the PV array. In addition, the multi-scale CNN performs better than the multilayer CNN when using the transformed features of all three GADF normalization strategies as input, mainly because the multi-scale CNN extracts features at different scales. When the CBAM module is embedded in the multilayer CNN, the introduction of channel attention mechanism and spatial attention mechanism enables the network to extract features of interest in the local area while focusing on the channel features, which is more selective in focus than the multi-scale CNN with different scales of convolutional kernels for feature extraction. Overall, the CNN with

CBAM module has the highest classification accuracy, with the stacked IV-PV graphical feature matrix transformed by the proposed Isc-Voc normalized GADF as the input. The diagnosis accuracies applied to PV arrays with two diode configurations are 99.62% and 99.40% respectively.

**Table 6** The fault diagnosis results of applying different CNN-based classification models.

|  |  |  | Testing accuracy | | |
|---|---|---|---|---|---|
|  |  | GADF | Normal normalization | Global normalization | Isc-Voc normalization |
| Circumstance 1: With blocking diodes | Multilayer CNN | IV | 98.46% | 98.25% | 98.85% |
|  |  | PV | 95.33% | 91.54% | 97.83% |
|  |  | IV-PV | 98.58% | 98.19% | **99.10%** |
|  | Multi-scale CNN | IV | 98.96% | 98.75% | 99.21% |
|  |  | PV | 96.75% | 91.44% | 98.81% |
|  |  | IV-PV | 99.10% | 98.65% | **99.31%** |
|  | Proposed CBAM-CNN | IV | 99.10% | 99.02% | 99.58% |
|  |  | PV | 98.35% | 95.83% | 99.08% |
|  |  | IV-PV | 99.42% | 98.98% | **99.62%** |
| Circumstance 2: Without blocking diodes | Multilayer CNN | IV | 98.50% | 98.04% | 98.42% |
|  |  | PV | 92.06% | 91.23% | 97.25% |
|  |  | IV-PV | 98.48% | 97.79% | **98.65%** |
|  | Multi-scale CNN | IV | 98.60% | 98.96% | 98.73% |
|  |  | PV | 94.21% | 92.92% | 98.33% |
|  |  | IV-PV | 98.85% | 98.54% | **99.06%** |
|  | Proposed CBAM-CNN | IV | 98.92% | 99.02% | 99.32% |
|  |  | PV | 96.06% | 94.10% | 98.81% |
|  |  | IV-PV | 99.15% | 98.77% | **99.40%** |

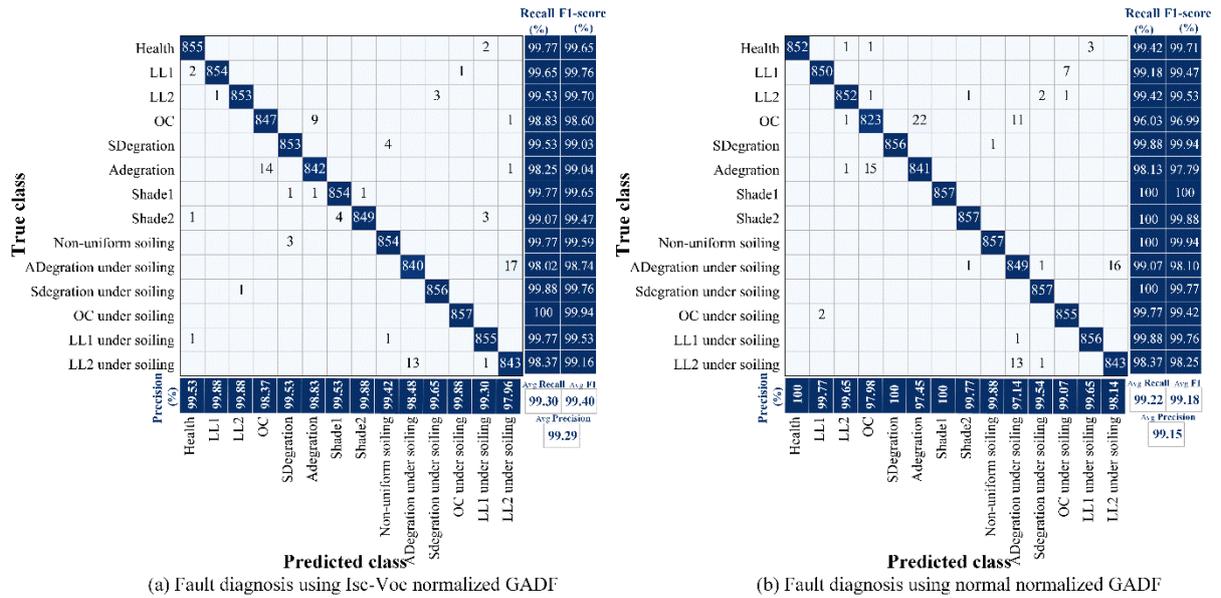

**Fig. 20.** The confusion matrices of PV arrays with blocking diodes under soiling condition.

Furthermore, the preprocessing methods of Isc-Voc normalized GADF and normal normalized GADF, which are both highly accurate in using the CNN-CBAM as classifier, are compared. Fig. 20 and Fig. 21 are the confusion matrices of fault diagnosis for PV arrays with and without blocking diodes, respectively. For arrays with blocking diodes, the two normalization strategies are similar in the discrimination ability of most fault types. The specific difference is that the Isc-Voc normalized GADF performs slightly better than the normal normalized GADF in terms of OC, Adegration, and Adegation under soiling, due to the similarity of these characteristic curves under some environmental conditions and levels of faults. The former reduces the proportion of OC misdiagnosed as Adegration by 59% (13 samples) and does not produce the 11 samples for which the latter diagnoses OC as Adegration under soiling. In addition, for arrays without blocking diodes, the characteristic curves of LL2 generated by current backflow are highly similar to those of Adegration and OC, etc. The Isc-Voc normalized GADF has higher discrimination than the normal normalized GADF in this case, and their overall recall and F1-score both are 99.28% and 98.82, respectively. Specifically, the Isc-Voc normalized GADF can avoid 11 samples of LL2 misclassified as Adegration, 15 samples of OC misidentified as Adegration under soiling, and more 11 samples of LL1 and OC under soiling that are not correctly distinguished. Moreover, the Isc-Voc normalized GADF is 27% more accurate than the normal normalized GADF in distinguishing between Adegration under soiling and LL2 under soiling and improved the diagnostic accuracy of complex faults affected by contamination by 41.94%.

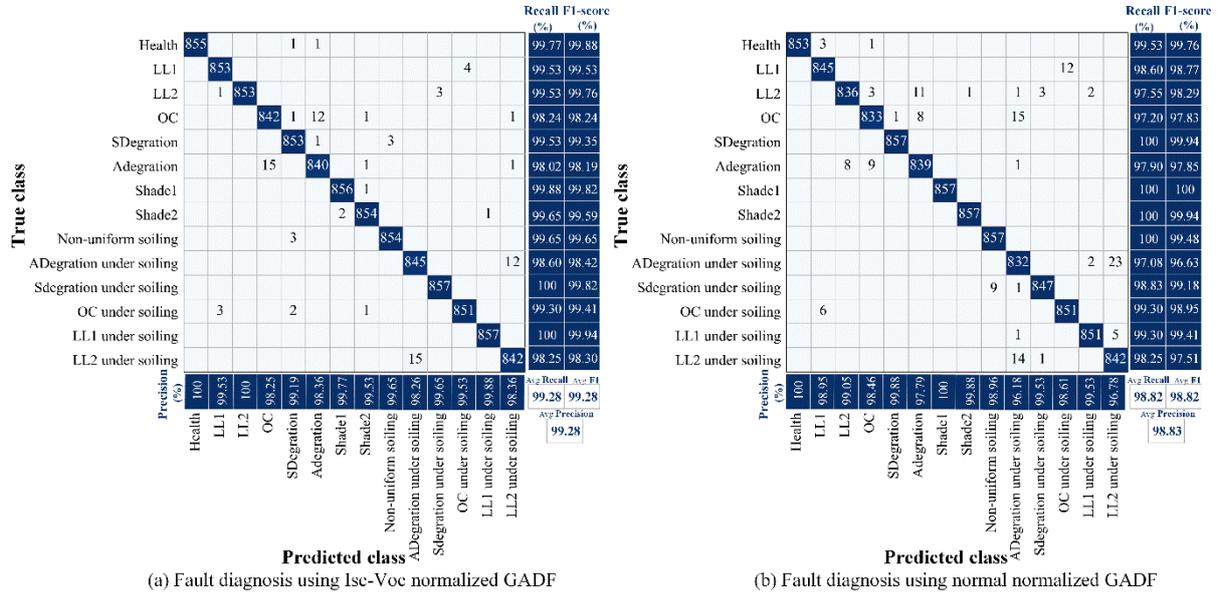

(a) Fault diagnosis using Isc-Voc normalized GADF  (b) Fault diagnosis using normal normalized GADF

**Fig. 21.** The confusion matrices of PV arrays without blocking diodes under soiling condition.

Generally, the complexity of faults increases due to the impact of dust, there will be individual errors in the classification of all fault types. And the proposed Isc-Voc normalized GADF method has higher accuracy in differentiating complex faults and is able to converge quickly during training, and the processed features are beneficial for the rapid learning of parameters, which provide the best classification performance when applied to the proposed CNN-CBAM, as shown in Fig. 22.

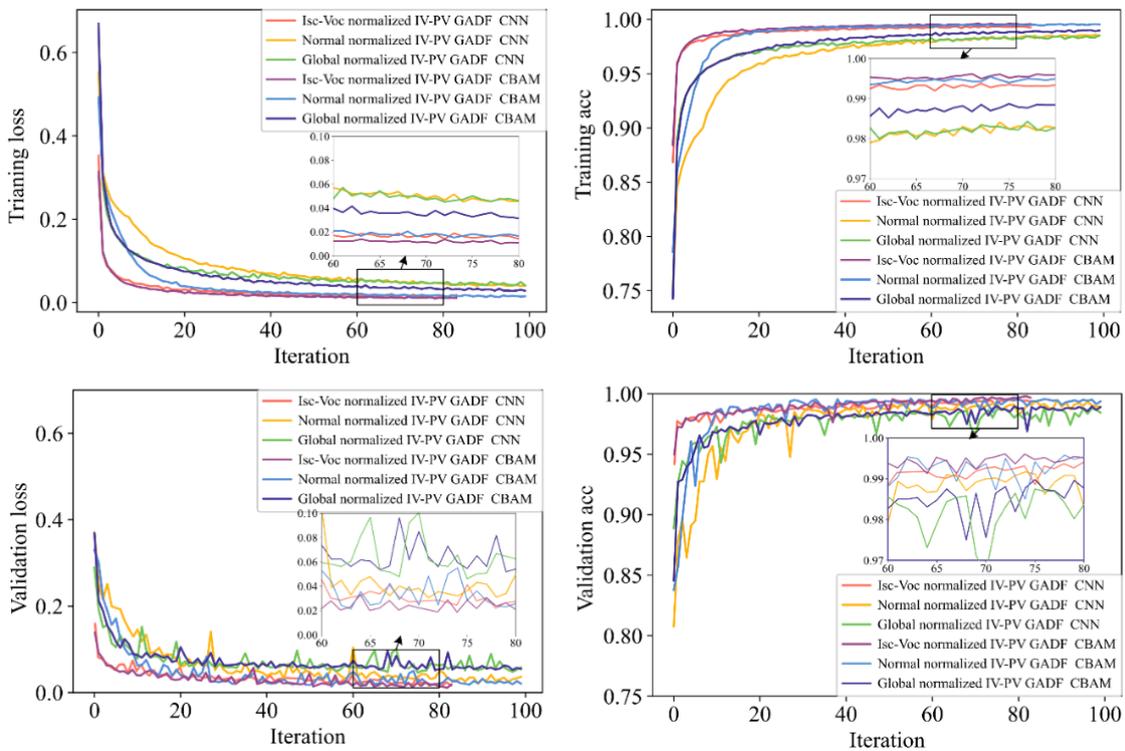

**Fig. 22.** The loss and accuracy for typical CNN-based classifiers with three GADF normalization methods.

### 4.4 Case 2. PV arrays under operating condition without non-uniform soiling impact

To further illustrate the universality of the proposed method, the performance of fault diagnosis under normal operating condition without dust impact is analyzed, which contains 9 fault states. Similarly, we analyze the influence of transformed graphical features processed by various GADF normalization methods as input on the accuracy of classification algorithms. Specifically, as shown in Table 7, for simple classifiers such as ANN, using the matrix of I-V or P-V curve converted by Isc-Voc normalized GADF as input results in a significant classification accuracy improvement of 97.57% compared to 93.26% for normal normalized GADF and 93.20% for global normalized GADF. And the accuracy of the proposed normalized GADF is up to 98.26% when the IV-PV stacked transformation matrices are used as the input features. Moreover, the CBAM-CNN exhibits general merit when the features processed by the optimal GADF strategy, i.e. stacked IV-PV transformed by Isc-Voc normalization, are applied to the CNN-based models. This is related to the fact that the types of faults to be distinguished in operation are not diverse and their complexity is general. In other words, for the faults not affected by dust, the graphical feature matrices transformed by Isc-Voc normalized GADF applied to a relatively simple structured CNN can obtain satisfactory results, but it is undeniable that the CNN embedded with CBAM module still has a slight accuracy advantage.

**Table 7** The fault diagnosis results of applying different classifiers and GADF methods.

|  |  |  | Testing accuracy | | |
|---|---|---|---|---|---|
|  |  | GADF | Normal normalization | Global normalization | Isc-Voc normalization |
| Circumstance 1: With blocking diodes | ANN | IV | 87.59% | 93.20% | 97.57% |
|  |  | PV | 93.26% | 91.02% | 97.57% |
|  |  | IV-PV | 98.09% | 93.68% | **98.26%** |
|  | Multilayer CNN | IV | 99.22% | 98.28% | 99.58% |
|  |  | PV | 98.41% | 96.99% | 99.38% |
|  |  | IV-PV | 99.42% | 98.54% | **99.58%** |
|  | Proposed CBAM-CNN | IV | 99.13% | 98.61% | 99.80% |
|  |  | PV | 98.83% | 97.57% | 99.51% |
|  |  | IV-PV | 99.18% | 98.35% | **99.84%** |
| Circumstance 2: Without blocking diodes | ANN | IV | 75.28% | 94.44% | 97.81% |
|  |  | PV | 82.74% | 91.34% | 97.89% |
|  |  | IV-PV | 94.44% | 94.83% | **98.22%** |
|  | Multilayer CNN | IV | 99.22% | 98.70% | 99.25% |
|  |  | PV | 98.39% | 96.95% | 99.35% |
|  |  | IV-PV | 99.22% | 98.61% | **99.42%** |
|  | Proposed CBAM-CNN | IV | 99.19% | 99.13% | 99.77% |
|  |  | PV | 98.51% | 97.54% | 99.45% |
|  |  | IV-PV | 99.25% | 98.38% | **99.81%** |

Fig. 23 and Fig. 24 show the fault classification results of Isc-Voc normalized GADF and normal normalized GADF for PV array with the two blocking diode configurations. It can be seen that the confusion matrix of the former does not contain many sporadic misclassifications of the latter and is expressed in the overall recall and F1-score of 99.81% and 99.18%, respectively. In the fault diagnosis of PV array with blocking diode configuration, the classification error between SDegration and Non-uniform soiling is 84% less in the former than in the latter, with 8 and 50 samples respectively, and the proportion is 80.43% in the array without blocking diode. This may be due to the low fault levels of these two fault types with dynamic fault parameters, i.e. the low fault differentiation between some SDegartions with low fault levels and low degrees of dust shielding. In fact, the correct classification accuracy for both SDegration and Non-uniformed soiling, which may be misclassified, still reaches over 99% for either diode configuration of array. Overall, the proposed CBAM-CNN model based on Isc-Voc normalized GADF transformation method is able to achieve high accuracy in faults classification, which is also applicable to ideal operating condition without dust impact.

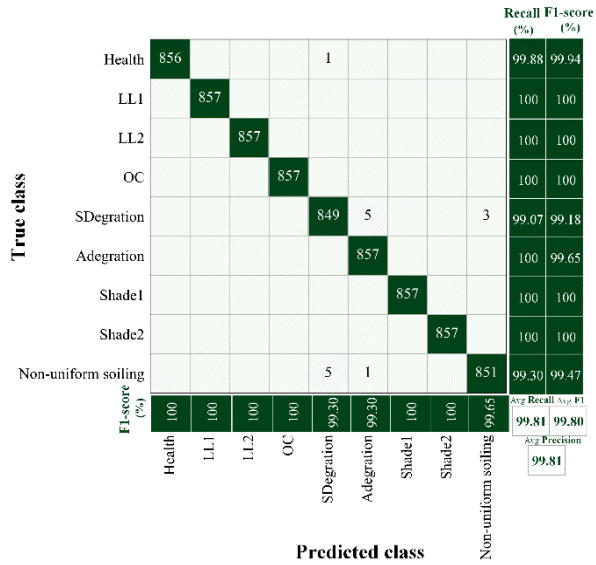
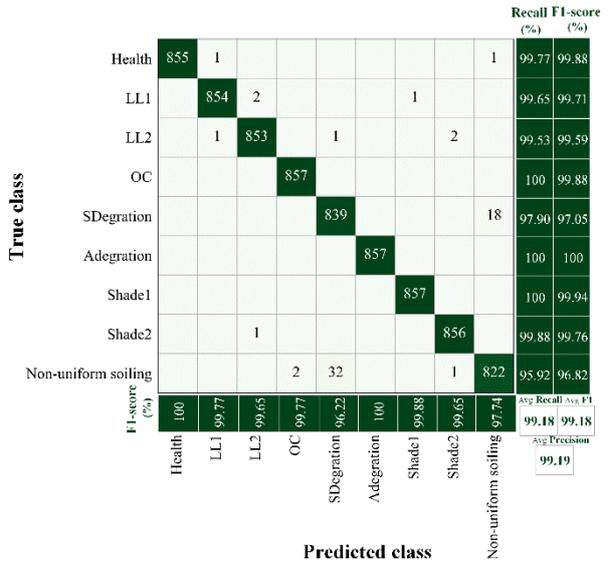

Fig. 23. The confusion matrix of PV arrays with blocking diodes.

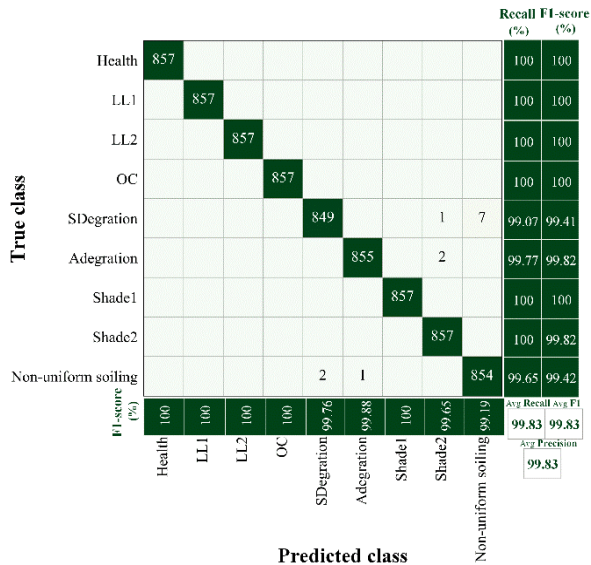
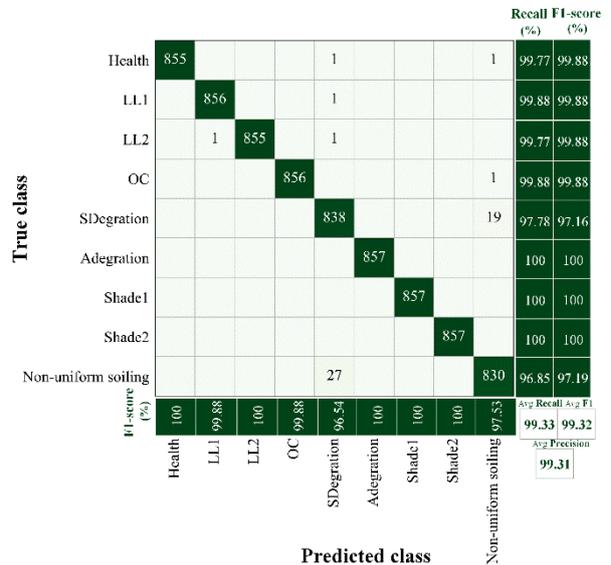

Fig. 24. The confusion matrix of PV arrays without blocking diodes.

The training and validation process of two blocking diode configurations are shown in Fig. 25 and Fig. 26. When the inputs are graphical features processed by Isc-Voc normalized GADF, the loss of model decreases the fastest and the accuracy rises rapidly compared to other normalization strategies, whether for ANN, CNN, or CNN-CBAM as the classifiers. It is consistent with the original intention that the graphical feature transformation method proposed can achieve high discrimination between similar faults in different environments. In particular, for arrays without blocking diode configurations, the model that classifies similar fault types using the transformed graphical features by normal normalization strategy requires a relatively long stage of learning, and the model is not as stable as models using other strategies. This is owing to the fact that the classification accuracy relies heavily on the feature learning capability of the classifiers when the discriminative ability of features is constrained. Moreover, in terms of the classification ability of models with different levels of complexity, simple models such as ANN require longer training time than CNN-based models, and the proposed Isc-Voc normalized GADF can significantly improve the diagnosis performance when applied to category of simple classifiers. For faults not affected by dust, the CNN-CBAM still has slightly better accuracy and training performance than multilayer CNN, although their differences are not significant due to the excellent strength of the CNN-based models in dealing with less complicated tasks.

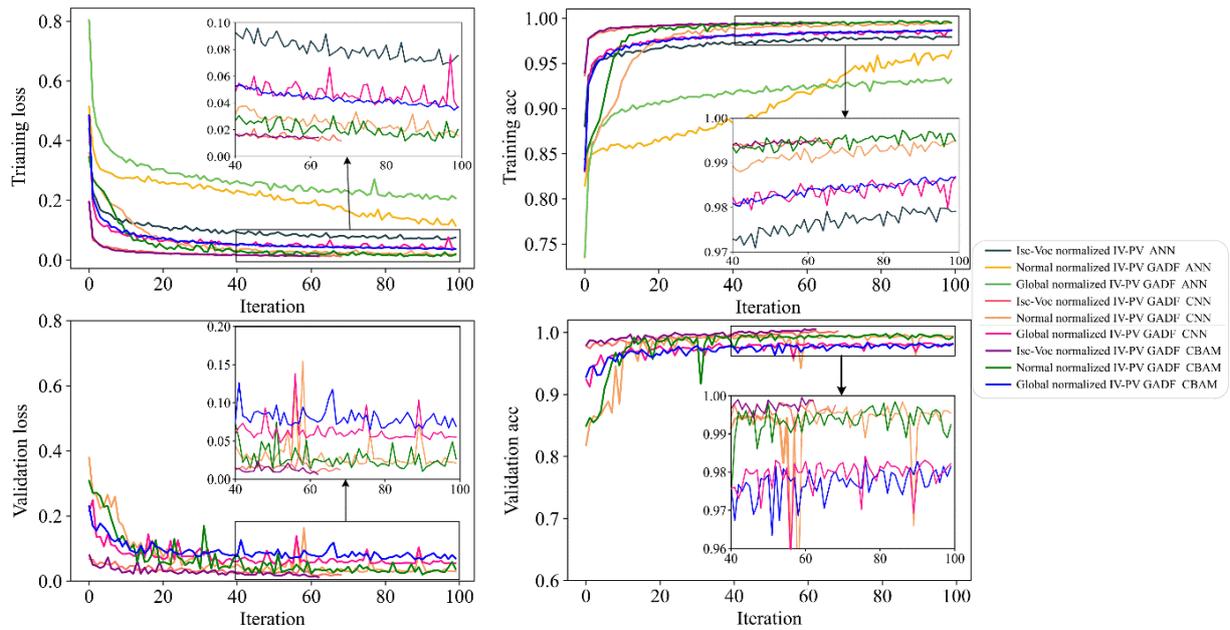

**Fig. 25.** The loss and accuracy of PV arrays with blocking diodes.

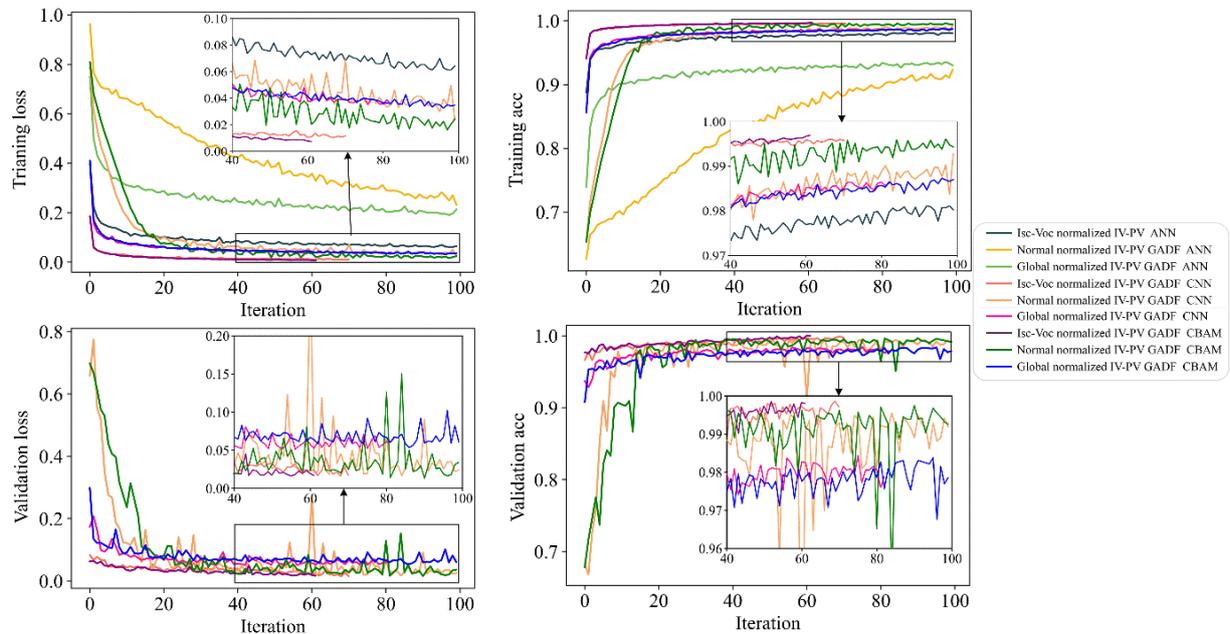

**Fig. 26.** The loss and accuracy of PV arrays without blocking diodes.

### 4.5 Discussion

Compared with the most advanced research in the existing literature, our approach enables the accurate identification of dynamic fault types for arrays with multiple scenarios and configurations. Among the categories using GADF transformed graphical features as classification features, the IV-PV stacked matrix by the proposed Isc-Voc normalized GADF preprocessing method outperforms the transformed feature of single characteristic curve in [39], both in terms of classification accuracy and performance of model training. This implies that the stacked GADF graphical features can highlight the distinguishability of different fault classes in preprocessing methods without employing correction of field characteristic curves, and that the GADF based on Isc-Voc normalization has a more reliable discriminatory capability. The main reasons for this are that stacked graphical features contain richer fault information than features of single curve, and that the Isc-Voc normalized GADF overcomes the challenge of inconsistent feature characterization of the same faults in different environments. In the comparative analysis of different classifiers, the proposed CNN-

CBAM has a certain accuracy advantage over other CNN-based methods, and is obviously superior to simple models such as ANN, especially under dust influenced operating condition containing multiple mixed and high-complexity dynamic faults. This is mainly because the characteristic curves jointly affected by dynamic fault parameters and changing environmental factors are more complex, and the learning ability of simple classifiers is limited. Indeed, it is worth noting that input features have a greater impact on classification performance, both in terms of accuracy and robustness, than the type of classifiers. The notion has been fully proven in terms of the significantly better diagnostic performance of Isc-Voc normalized GADF features compared to other GADF features using simple model ANN as the classifier. Furthermore, the benefits of the proposed Isc-Voc normalized GADF are still notable and effective in the condition of not considering the impact of dust, and CNN-CBAM has slight privileges over the basic CNN models due to the general complexity of faults in this scenario.

These results prove that features are extremely important, that is, the original data converted into effective features can improve the discriminatory quality of the input features. It can simplify the tuning process of the classifier and improve the diagnostic performance. Moreover, the CNN structure with the introduction of the CBAM attention module, based on high-quality effective features, offers more significant advantages in dealing with the diagnosis of complex fault types.

## 5. Conclusion

In this paper, we propose a new fault diagnosis method for identifying and evaluating faults in PV arrays, which is able to cope with PV arrays with different blocking diode configurations under both operating conditions considering and not considering the dust impact. It is summarized that the salient aspects of our approach are as follows. To facilitate the extraction of the full information in the characteristic curves by the classifier, we transform both the I-V and P-V curves into a graphical feature matrix of GADF and stack them together. Then, for reliable and easy-to-implement practical applications, which require that features of the same faults in different environments are effectively uniform, we adopt calibration-independent field characteristic curves and normalize them by GADF using the ideal Isc and Voc under their ambient conditions. Furthermore, to overcome the challenge of classifiers for multiple dynamic faults in complex operating conditions, we employ the convolutional neural network embedded with CBAM modules to identify and classify the graphical features of complicated faults.

The performance of the proposed technique is validated in terms of accuracy and stability in various scenarios. Specifically, the Isc-Voc normalized GADF transformed graphical features have higher fault discrimination than the features transformed by the other strategies, significantly simplifying the processing of the classifier and obtaining magnificent diagnostic accuracy for both conditions with or without dust effects. And it is verified that CNN-CBAM is effective for the identification of normal operating condition containing relatively simple faults or dust operating condition containing complex faults. In general, the proposed method based on GADF-transformed graphical features of characteristic curves and convolutional neural network with CBAM modules is applicable to PV arrays with or without blocking diodes and is also effective for fault diagnosis of different operating conditions, including concurrent faults affected by dust, which has economic benefits. And the proposed Isc-Voc normalized GADF transformation provides a new scheme for using full characteristic curve information with discarding STC correction and relieves the experimental dependence on correction factors, which greatly expands the universal application of fault diagnosis. Based on this research, there are still challenges that need to be addressed in future work, such as the further differentiation of different dynamic fault levels and the location of strings occurring faults while diagnosing, which will play an important role in studying the improvement of O&M efficiency and return-on-investment.

# Acknowledgement

This work is supported by the National Natural Science Foundation of China (No. 61573046) and Program for Changjiang Scholars and Innovative Research Team in University (No. IRT1203).